# Study of the fundamental physical principles in atmospheric modeling based on identification of atmosphere - climate control factors

### Version 1.0 (July 2010)

### PART 3

**ENFORCED DEVELOPMENT OF THE EARTH'S ATMOSPHERE**

**PHYSICAL AND TRANSCENDENTAL DIVISIONS**

**PHYSICAL DIVISION: OZONE-OXYGEN TRANSFORMATION**

**IN THE EARLY ATMOSPHERE**

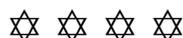

## TABLE OF CONTENTS

# ABSTRACT

**Index: Control factors of the Earth's atmosphere-climate system, active life forcing in physical and transcendental divisions, example of the Arctic bromine explosion, modeling of natural phenomena**

This part of a study is a continuation to the previously started discussion of the control factors of the Earth's atmosphere-climate system ([10,11]). We review some basic issues of the life-prescribed development of the Earth's system and the Earth's atmosphere and discourse the unity of Earth's type of life.

Earth's type of life and Earth's system operate in two divisions: physical and transcendental. Considering the enforced development of atmospheric compartment of the Earth's system, we reconstruct a legitimate relationship between these two divisions. Situated in physical division, Earth's life forms are spirited up by transcendental life. Provided by transcendental life, liveliness of the biotic Earth is expressed in (1) the self-reflecting and enforceable spiritual component of sensing and (2) intentional activities of the Earth's life forms. Yielded life, impact of the Earth's life forms on the energy-matter transformations and development of physical reality is possible due to the temporal <probabilistic> nature of the Earth's material phenomena.

In everyday-world division, Earth's life forms capture and fulfill enforcement through their active sensing (partaking and observation) of material phenomena. To start with, we exemplify a metabolic ozone-oxygen sensitivity in the Earth's surface-dwelling biota. Using a conception of oxygen biogeochemical cycling, we substantiate the origin of atmospheric phenomena in the metabolic pathways acquired by the Earth's type of life.

Atmospheric phenomena can be explained in terms of energy transformations. We are especially concerned with elaboration of the energy transformations. For instance, elaboration of the atmospheric ozone-oxygen transformations has existed since the time of the gradual spread of the oxygen photosynthetic microbiota (microbial organisms) and its food webs over the planetary surface. We provide an analysis of the present-day atmospheric processing, namely the coupling phenomena of elaborated ozone-oxygen transformation and Arctic bromine explosion ([10,11]).

We point out that modeling of natural phenomena, e.g. atmospheric processing, must include the explicit expression of the active life forcing distributed all over the physical division of the Earth's system. We commence this paper by detaching sensing phenomena in a human observer into physical (Earth's life) and transcendental (life) divisions, and suggest a variety of propositions which should be true for all Earth's type observers of material phenomena.

The asserted reexamination of physical reality within the Earth's system signals return to the long-time traditions of natural philosophy. The paper is suitable for the popular reading.

## § 1 JUSTIFICATION OF RESEARCH

**Index: Purpose of the Earth's development, successive elaboration of the energy transformations, microbiotic metabolism, adjustment of metabolic emissions, oxygen biogeochemical cycling, oxygen-ozone transformation, multiple unity of the Earth's operations, extreme environments**

We are motivated to demonstrate a broad perception of the biotic Earth's system and to employ this perception in the study of the atmospheric phenomena. Broad perception seems to be able to provide the remarkable clarification in regard to many of the scientific inquiries about the physical reality. The reexamination of physical reality within the Earth's system signals revival of metaphysical discourse on the study of material phenomena. The very meaning of this discourse is a requirement of universal understanding and deep knowledge of physics, philosophy, chemistry, biology and mathematics, etc.

Since civilized life style brought about deficiency of impressions and sequential decrease in the irrational human sensibility, there are certain concerns about the easiness of acceptance of the conception of transcendental life. In fact, as a result of deficiency and dogmatization, the mainstream scientific research and teaching of the last centuries moved to the methodology entirely different from the methodology of natural philosophy and created a machinery image of a dispirit and purposeless nature. "People become so used to that they lack the realization that anything is missing" (I. Asimov," Star dust").
As we put it together, the meaning of this research is in search and exertion of the human civilized life form for the sake of pursuit of a higher existence.

In metaphysical space, the final purpose of the Earth's development is development of the Earth's type of life into the highest form of itself. We suggest that the highest form of the surface-dwelling Earth's life will be the one featuring the elaborated applications of the external energy sources, and foremost, the incoming space radiation. We describe several basic measures of elaboration that could be determined from the observation of superficial habitat. These measures – connectivity, access control, sensitivity dependence and waste utilization in the Earth's system, are the most important offshoots of energy elaboration at the level of the Earth's life as a whole.

Development of any physical system is accompanying by expansion. As far as we are concerned, life has expanded its control over the planet. One time in a past, life and the Earth's life took the full control over the atmosphere by adjusting emissions to the energy inflow from (i) the outer space and (ii) the inner Earth. The most familiar example of adjustments is an oxygen emissions' adjustment. From the thermodynamic considerations, such adjustment compelled existence of the surface regions with liquid or solid water , which means it compelled low temperature conditions. As it happens today, emitted oxygen was going to the early atmosphere for the chemical cleansing and transformations, and then had been returned to the living organisms and chemical sinks such as reduced geothermal outflows and rock. Starting the early stage of development of the biotic Earth, oxygen by-product of oxygenic photosynthesis has been a countable part of the total oxygen emitted into atmosphere. Being very reactive, free oxygen (ozone and oxygen) in high concentrations is toxic for the Earth's type of life. Through their metabolic pathways, Earth's organisms successfully offset toxic concentrations of free oxygen in physiological medium and environment, e.g. the Earth's atmosphere.
While atmosphere plays a role of a reservoir of the metabolic by-products which have the fastest turnover in biosphere, recycling and redistribution of chemical species over the atmosphere are subject to the input of (1) space energy and (2) heat of the inner Earth.

Life regulates the development of the Earth's planetary system. Multiplicity of life-favorable environments on the Earth's surface came into being as a consequence of the biogeochemical development. In our study, the habitable environments that we speak about are located on the Earth's surface.

Present-day biogeochemical cycling, atmospheric and other global scale processes reflect on their foundations. At the present stage of development, superficial conditions allow a wide spread of the oxygen photosynthetic microbiota and its food webs comprising the large life forms. Discussing the aspects of oxygen biogeochemical cycling, we accentuate ozone-oxygen conversion as essential for the gradual spread and survival of oxygen photosynthetic organisms over the Earth's surface.

Early biotic phenomena are better grasped in extreme environments. It had been resumed that it is microbiota that creates a multiplicity of optimal habitable environments by the redirection of energy and matter cycling inside the Earth's system ([16], [11]).

Because of the high efficiency of the oxygen primary reactions, oxygenic photosynthesis upon visual light in the surface-dwelling biota grew into a main developmental issue for the superficial organic life. Oxygen primary reactions (oxidation-reduction reactions) involve molecular $O_2$ as an electron acceptor. These primary reactions are efficient both in physiological medium of organic matter and in atmospheric medium. In atmosphere, among others, oxygen primary reactions include (i) catalytic ozone-oxygen conversion and (ii) formation of hydroxyl radicals out of water and oxygen. During the atmospheric recycling of free and combined oxygen, oxidation of metabolic by-products leads to production of oxide species. Whether by photochemical or electromagnetic (e.g. lightning) processes, some part of the atmospheric $O_2$ is converted to ozone $O_3$. If molecule of ozone comes in contact, it destroys organic cell by disintegrating its transmembrane protein (lipids, fatty acids). Contact with ozone could bring deadly consequences for an unicellular organism.

Control over the desirable levels of surface ozone is carried through influx of trace volatiles into environment: for instance, ozone decomposition is catalyzed by many biological metabolites and their derivatives. Surface and near-surface degradations of ozone prevent damage to the surface-dwelling organisms and intensify oxygen flow into biosphere. We infer that atmospheric ozone-oxygen conversion and destruction of the surface ozone had to exist prior to the advent of the global oxygenic photosynthesis.

As known, sunlight condition alone is not sufficient for the transition to the productive season in oxygen photosynthetic microbiota. Happening in polar sunlit conditions, massive surface ozone-oxygen transformation conveys the Arctic "seasonal transition" message to the surface-dwelling organisms of the Northern Hemisphere and, consequently, unlocks their metabolism ([11]).

Though early microbiotic phenomena are distributed over the Earth's system, they are more easily observed in extreme environments. For instance, in extreme surface environments limited in the solar light energy, the early phenomena can be qualitatively accessed through the atmospheric chemistry and photochemistry of metabolic derivatives. Implying the interrelation of biogeochemical cycles of terminal oxidants in the extreme environments, apropos for an assessment species are: sulfur and oxygen species, halogen and oxygen species, nitrogen and oxygen species, etc.

Pro-life conditions in extreme environments is the effect of choice rather than chance. The first and foremost, extreme environments correspond to the microbiotic metabolism; they secure conservation, existence and development of paleo metabolic pathways. Survival of paleo microbiota < paleo metabolic pathways> is essential for the interminable development of the Earth's system.

In our explanations about the foundations of biogeochemical cycling, we extensively use polar Arctic phenomena. Beforehand, in [10] and [11] we have discussed properties of the tropospheric phenomena of the natural bromine emissions at the polar Arctic sunrise. Associated phenomena is an interception of bromine and oxygen biogeochemical cycles. The phenomena reveal information about the underlying metabolic schemes in the surface-dwelling microbiota. Via ocean bromine concentrating, surface bromine emissions to the Arctic atmospheric boundary layer (ABL) can be indeed traced to the paleo marine microbiota.

## § 2 INTRODUCTION TO DEVELOPMENT OF THE BIOTIC EARTH
**Index: External energy sources, solar-terrestrial relations, energy exchange by means of interplanetary magnetic field, microbiotic control factor vs. solar magnetosphere and radiation factor**

Our current assertion of the main energy sources of the biotic Earth is biased on the human experience. Therefore, our current assertion is ambivalent and inconsistent. Sequel of the inconsistency and of quantitative reductionism, scientific evaluations of the Earth's energy budget often tend to ignore heat flow leaking from the Earth's interior ([26]), tidal and other incoming energy fluxes and assume the solar energy source exclusively. Yet implying the dialectic divergence <ontogenetic dichotomy, duality>, the energy and biogeochemical budgets require a consideration of the numerous < at least, a pair of> sources. In truth, Earth's system is a biogeochemical system which takes (i) the most of its energy from two vast sources: (1) the Earth's molten core and (2) the Sun; and (ii) the most of the building matter from the inner Earth.

A great deal of understanding is that the Earth's energy-matter sources are straightforward regulated by life and the Earth's life. The connection is anything but obvious.
We make a notion of the obvious connection between (i) sea-floor spreading and subduction zones from one side and (ii) the emergence of the paleo marine microbiota activities from the other side. We also make a general notion of the cold microbiota and its food webs on the Earth's surface that provide recreation of the Earth's troposphere and pro-life conditions in the marine habitable environments in the interests of the Earth's life. Existence of the organic Earth's type of life relies on the existence and certain parameterization of the Earth's hydrosphere. Chemical, physical and dynamical recreation processes in troposphere support three states of water on the Earth's surface.

Long- and short-term varying solar activities are responsible for the variations of the energy fluxes between compartments of the Earth's system. Current knowledge of the solar factor variability and its various effects on the Earth's biota and, hence, its effects on the terrestrial climate is far from sufficient. It must be admitted that present-day solar activities are very much pro-life activities. The question is arisen whether there is a bi-directional connection between solar and biotic Earth's activities. Because of many significant cycling relations by which Sun and Earth relate to each other, analysis of the solar-terrestrial processing is incredibly difficult. The hard part of understanding how solar activities and the Earth's life are connected is that the Sun and the Earth are energetically connected through their magnetospheres. The variations of the interplanetary magnetic field (IMF) are recognized as far more important solar energy factor than increased velocity or pressure of the solar wind, sunspot numbers, etc. Up now, science and technology have allowed to learn the energy-information exchange between magnetospheres only as one way exchange - the decreeing Earth's response to the solar activities.

Variations of IMF polar slants support the Dungey's idea of interconnecting magnetic field lines and an open magnetosphere. A key way of energy exchange between magnetospheres is a magnetic reconnection. A magnetic reconnection is associated with the a world-wide magnetic disturbances. While the world-wide scale disturbances near the Earth's surface are small (50-300 nanoTesla out of a total intensity of 30-60,000 nanoTesla; phenomenology: a steep decrease of the field over 6-24 hours, then gradual recovery throughout 1- 4 days), the disturbances in the Van Allen radiation belts are large. Magnetosphere's disturbances present the observed radiative, chemical, and dynamical forcing for the Earth's atmosphere and Earth's superficial and beneath surface environments.
The Van Allen radiation belts are a pair of doughnut-like rings of plasma (ionized gas) trapped in orbit around Earth; plasmatic flows respond to a variety of solar events. The energy exchange between magnetospheres proceeds through a magnetic reconnection of plasma. At the time of coronal mass ejections and flares on the Sun, measured electric currents in the ionosphere about 3-4 times the norm. The leading edge of the magnetosphere, which usually sits at 76,000 km from Earth toward the Sun, is pushed in to 25,000 km (adapted from NASA website *http://www-istp.gsfc.nasa.gov/istp/news/9812/radiationbelts.html*). The informational exchange between magnetospheres is waiting for its exploration.

Although the role of life and the Earth's life in processing of these powerful energy exchanges is yet to be understood, it is verified that energy exchanges partly proceed over the atmospheric medium. Earth's life controls the development of atmosphere-climate system.
On geological timescales, climate variations are consistent with solar activities. On the smaller timescales, from 1 day to around 10, 000 years, the Earth's system modifies atmospheric properties (like aerosol loading and humidity, cloud cover) toward the spectral irradiance variations that benefit the Earth's life and, especially, surface-dwelling marine microbiota. These modification poses new unresolved constraints on the atmosphere-climate system. Incorporated Earth's life, the biogeochemical cycling effects atmospheric radiative, chemical and dynamical changes which are subsequently transmitted to the magnetosphere.

Similarly to the IMF disturbances which stress the Earth's system from the radiation belts to the superficial and under earth habitations during the solar stressful events, the superficial environments and the lower atmospheric layers influence on the radiation belts. It was found that equatorial ionospheric densities are modulated by atmospheric waves driven by the persistent tropical rainstorms: lightning strokes during the rainstorms generate radio waves which interact with <radiation belt> plasma to clear a safe zone between the inner and the outer belts. Gravity waves also produce stress on the ionosphere: at equatorial latitudes, gravity waves generated by hurricanes and typhoons may seed plasma bubbles in the ionosphere.

It is fairly possible that the crust's motion and magnetic anomalies are a result of the interminable microbiotic activities thoroughly the development of biotic Earth. Heat leaking from the Earth's core is necessary for some microbiotic activities. In under earth environments, for example, on the deep sea floor, microbiota activities regulate the global chemistry of the crust and ocean. Sulfur is a terminal oxidant of deep ocean microbiota, and hydrogen sulfide is an end product of oxidation of marine organic matter. Because of sulfur and iron chemistry coupling, the biogeochemical fluxes of sulfur species from and into the Earth's crust - ocean barrier impact magnetization of the oceanic crust.
Relevance of magnetosphere's changes from the one side and cycling of the Earth's orbital inclination (80,000-100,000 yrs) and wobbling of the Earth's tilt (26,000 yrs) upon the tidal forcing of moving celestial bodies from the other side is also significantly impacted by the Earth's life control factor over a body of the World Ocean.

We have not invented here an entirely new reason, there were the Babylonians who said that the Sun-god provided laws <technologies of energy transformation> for people to follow. Hopefully, in the future years we shall find out how to correspond solar activities and the Earth's life control factor. It shall require a transcendent knowledge and a mathematical projection of higher dimensions to the 3+1 everyday-world to formulate origins of the solar-terrestrial relations. In the meanwhile, since solar-terrestrial relations are very complicated, they are not really presented in the solar dynamo models. Dynamo models of independent solar activities range from simple, strictly periodic, deterministic models to the stochastic models: Parker's migratory dynamo, Solar Parker spiral, Pikovsky's stochastic resonance in a magnetic dynamo model with a pure Poisson occurrence rate for grand minima ([21],[23],[24] and [34]).

## § 2.1. EMERGENCE OF PRO-LIFE SUPERFICIAL ENVIRONMENTS
**Index: Co-integrated activities of the cold and hot microbiota, food webs, human civilization as a part of the oxygen photosynthetic food web, energy-matter-information equivalence**

The emergence of the paleo marine microbiota had represented a start of the surface-dwelling Earth's biota. There is no commentary on the questions (1) whether, how and why the Earth's life was first introduced to the innermost of the planet and (2) at what set of conditions the Earth's life had occupied the planetary surface. However, we are able to state that arrival of the surface-dwelling biota had required cool conditions at least at the part of the planetary surface. The other necessary condition was accessibility of liquid water. Paleo marine microbiota is, probably, the first superficial form of the Earth's type of life and a prime superficial contributor to the intensive biotic emissions of free oxygen into the Earth's atmosphere. The next necessary condition is availability of solar radiation, especially, solar radiation in visual and infrared spectrum, which determines production for both aerobic and anaerobic superficial microbiota.

Metabolism of oxygen photosynthetic food webs also include the oxygen consumption schemes. Rates of oxygen production and consumption for the surface-dwelling organic life depend on the pro-life conditions of habitat (ozone and oxygen concentrations, temperature profiles, etc.). Emergent life-favorable environments on the Earth's surface are a result of the biogeochemical Earth's development upon life, Earth's life and external energy control factors. First pro-life environments were built by the co-integrated activities of the cold and hot microbial organisms.

The development of the Earth's system had led to creation of the divergent food webs above the primordial microbiota. Already on the base levels of hierarchy, food webs are co-integrated with each other through feeding and waste utilization.

On Earth, superficial biota is associated with liquid water which can exist only in specific ranges of temperature and pressure. Arguably, paleo marine microorganisms became first superficial form of the Earth's type of life and a prime contributor to the development of the stable oxygen-rich Earth's atmosphere (free oxygen content of 10% - 35 %). At the present state of development, oxygen content of ABL is about 15%-21%; superficial conditions allow the wide spread of the oxygen photosynthetic microbiota and its food webs, which include large life forms.

We assert that human life forms belong to the food web of the oxygen photosynthetic microbiota. Modern anthropogenic activities are characterized by the very intensive energy transformations in ABL, and, particularly, in ABL of the Northern Hemisphere. Human civilized life form has tendency to increase energy consumption; it is in a continuous search for the new energy sources and available energy transformations for the applications of light energy (and information) in social, industrial and agricultural activities. The increased energy consumption occurs independently of the energy transformations required by human physiology; present-day energy consumption and transformations for the needs of human civilization don't take a full advantage on the information-energy equivalence. Nearly all man-made applications are not suited to the requirement of effectiveness of utilization and elaboration.

Energy-matter-information equivalence is a constitutional amendment to sensing at the base levels of the Earth's life. Active transcendental life component is distributed among a mindset of multiple agents, purposeful and competent on their own levels of hierarchy. Active life agents (organisms, populations, life forms, etc) transform external energy and recycle matter. By means of the changing heat and matter gradients in the shared environment, agents create "knowledge" messages to the other active life agents. The knowledge effectiveness is determined by its amount, content quality and reliability (see example of Arctic bromine explosion in [10] and [11], amount and composition of Arctic ocean life emissions). Metabolic emissions into environment can be of the homeopathic dosage and still produce the significant temporal-spatial impact. Corresponded to the lifespan activities of a specific life form, biotic-related emissions affect the shared environment of a life form, hence forcing the other presented life form to act upon.

Information-energy equivalence has not been researched yet throughout scaling ranges of the biotic Earth's system. However, there exist some prerequisite methods, known as topological methods, which were first ramified for the astronomical and astrophysical research, are also applicable to the kinship phenomena of the Earth's system.

Earth's system has an enormous volume of knowledge <context> in the physical-chemical domain. Captured by active life component, organizational-context messages are being used to optimize multiple parallel operations. In contrast to the physiological operations of human organism, of its organs and its tissues which are optimized at no intervention of human rational perception, - industrial, agricultural and social activities of human population are strongly influenced by the human rational enterprise. The rational activities of civilized life form, namely anthropogenic factor, represent the hardship for the Earth's system; they are not optimized, they may redirect energy and matter in the "confused" way and, thus, they may create unreliable messages to the other life agents.

## § 2.2. SENSING AS A FOUNDATION OF LIFE
**Index: 4 levels of sensing, purpose of existence**

There is ample evidence that life-favorable environments vary from species to species and for many species also depend on the immediate physiological growth state of cell or cells. The formation of pro-life environments became available due to the microbiotic sensing. According to the latest findings of the cellular physiology, there is no fundamental difference between microbiota sensing and sensing in the large life forms. We differ between (1) an individual sensing of organism, as an instance of a life form, (2) a quorum sensing of population within a life form (or life forms), (3) a quorum sensing of a life form as a whole and (4) a quorum sensing of the Earth's life as a whole. Human experience and human perception provide some rational comprehension of the human sensing at the levels (1), (2), (3).

The transcendental (irrational, intuitive, etc) perception is most likely derived directly from (4). Transcendental perception, or spiritual experience, is given to the infinite rational interpretations different from one instance of a human life form to another (ancient Greeks, B. Spinoza, I. Kant, H. Helmholtz, H. Poincare, M. Plank and others). It also allows a multitude of interpretations in the same instance of a life form. Depend on an individual situation, some of these interpretations may turn into the powerful philosophical ideas or the heuristic scientific theories (Fractal theory, Gaia, Quantum mechanics, Chemical conception of the Ether, etc), striking music masterpieces, etc.

In human life forms, a quorum sensing of the whole of the Earth's life is actualized as an irrational perception of the Earth's beauty, as a wonder of the world's order and a wonder of life. At the beginning of the human civilization, the recognition of the life unity <in a form of faith, in an ideal frame for all interactions and reasons in the universe> was absolute. Later on, following the agriculture and urban development and establishment of primacy of rational <symbolic> reasoning, the recognition of a quorum sensing of the Earth's life as a whole became vague at least in the Western countries.

In a word, not only individuals, but human civilization should imply a quorum sensing (4) in its decision making in order to survive and to develop. Even it doesn't guarantee the survival of an individual organism or single population, a quorum sensing of the Earth's as a whole (4) is necessary for the survival of any Earth's life form. There are signs that human individuals regain a stronger quorum sensing at the tough times of necessity.

At all levels ( (1)-(4)), sensing is an instrument of the exploration of the physical reality. Sensing, perceiving and consciousness in the Earth's type of life conclude a self-reflecting emanating component. Because of this self-reflecting component, sensing, consciousness and perceiving could have been explored by the classical philosophical science (Plato's Metaphysical Epistemology, Leibniz' Discourse on Metaphysics, etc). Classical philosophical science holds a metaphysics view on the physical reality as being fundamentally purposeful. It is undeniable that exploration of the physical reality both by the individual and collective minds is ultimately dependent on the a-priori held expectations and values. For G. Leibniz, D. Mendeleyev, L. Boltzmann, M. Planck, A. Einstein and many other thinkers, the metaphysical beliefs were a key and a method for exploration of nature and mind. Advancing of science as a body of knowledge has been inseparable from research and enlightenment. Enlightenment appears to be sourced both in belief and in human reasoning and linked to the Earth's life as whole. In the words of Plato, "man is a being in search of meaning". We know many examples of the scientific discoveries that have been enlightened by faith <by a-priori logos, by metaphysics> and private enterprise towards a meaningful existence. From our experience, we learn that enlightenment is not completely dependent on the human rational reasoning. It also contains the probabilistic elements. Together with self-reflecting emanating ability, ability to attain enlightenment run beyond the human life form and beyond the Earth's type of life.

In essence, the future course of humankind as a part of the Earth's planetary system is determined by understanding of the purpose of existence. Purpose of existence can be learnt in numerous ways. It can be also learnt from the records of the Earth's development and from the analysis of contemporary natural phenomena.
Purpose for the Earth's life existence is a development of the Earth's type of life into the highest form of itself. Likely, the highest form of the surface-dwelling Earth's life must feature the diversity of the elaborated applications of light energy and information. It seems that mankind is in process of learning the natural (and universal life) principles of energy elaboration. Applications of light energy and information attributed to the paleo metabolic pathways are incorporated in the present-day operations of the Earth's type of life. The most powerful, controlling the Earth's energy balance and energy and matter distribution, these applications are exceptionally important for understanding of the natural principles of elaboration at the level of the Earth's life as a whole.

## § 2.3. SENSING AND FORMATION OF SUPERFICIAL BIOFILMS
**Index: Biological sensing, biofilm, optimal symbiotic environment**

On the biological level of comprehension, sensing could be translated into the concepts of physical science of matter. Most microbiota sense the environment by help and adaptation of a transmembrane protein and some by help of cytoplasmic receptors. Sensing in microbiota is an active, outward, part of cellular physiology. Generically, Earth's system microorganisms have sense of the nutrients and toxins, water, pH, temperature, and the magnetic field. In addition, the sensing includes a quorum sensing of the large populations that usually grow to the biofilms. As we noted already, quorum sensing (3)-(4) involves perception of non-human life forms and Earth's operations at global and micro scales; hence, quorum sensing is not falsifiable (testable) in framework of the modern earth and life sciences alone.

Biofilm formation is a logical aftermath of sensing in microbiota. Nearly all known microorganisms have mechanisms by which they can adhere to surfaces and to each other. Initially, a biofilm was understood as a thin structured layer of microorganisms, extracellular polymeric products and organic matter adhering to the solid-liquid surface. Large life forms owe their emergence to the biofilm formations, because integrability (or non-integrability) of a biofilm is directly related to growth and differentiation of tissue and organs of large organisms.

Extreme environments provoke the growth of biofilms, because biofilms are an important tool of microorganism's resistance to the harsh environment conditions. Biofilm co-existence of micro organisms and populations seems to be is a necessary condition for the divergence in the Earth's type of life. Biofilm presents a distinct stage in the development of microbiotic populations toward the symbiotic environment. In fact, biofilms allow the specification of microorganisms and stimulate symbiosis inside and between the food webs. For example, containing hydrocarbonoclastic bacteria (HCB), biofilms eliminate petroleum oil from the contaminated oceans and marine systems; or the soil biofilms, the superficial biofilms, maintain the stabilization and denitrification of soils; or anaerobic wastewaters biofilms of sulfate-reducing and methanogenic populations preserve the water quality. All biofilms feature the nutrient complementation in partaking populations, efficient waste recycling inside a biofilm and in its vicinity, and stable biomass accumulation while at homeostatic conditions.

There are more departed ways to understand the biofilm structure and operations. Biofilms could be seen as the surfaces of biotic-abiotic interactions*.
Sensing in microbiota activates the metabolic schemes, which correspond to the formation of the life-optimal environment. Under such an optimal environment we assume the sizable and expensive environment, optimal for the entropy production and population growth.
There are local pro-life environments which are optimal for the certain part of the Earth's type of life, and there is a global optimal environment representing the integrated purpose of the Earth's life. Analysis of the optimal global characteristics confirms that divergence and multiplicity in the Earth's type of life are required for the stable and gradual development of the Earth's system. Divergence and divergency (multiplicity, infinite series, *math.*) are the in-built property of the probabilistic <temporal> physical reality.

* Biotic-abiotic processes on the planet Earth are not separable by nature, because of physical design of the Earth's life:

   (i)     The biogeochemical cycling is processing over all possible spatial scales
   (ii)    Organic Earth's type of life lives on the external energy and matter sources. Already in microbiota, the metabolic processes have their input/output outside of physical boundaries of organism. Communication network of distributed active life components gets into the vicinity of a body of the Earth's type of life

In regard to our study, we propose that formation of the base-levels < pro-microbiotic> optimal symbiotic environment takes place within a biofilm and vicinity. Concept of a biofilm and vicinity can be extrapolated over the area and depth (Theory of Solution for Innovation Problems - TRIZ, extrapolation methods in innovations were studied by G. Altshuller, the originator of TRIZ). Upon this view, the atmospheric compartment of the Earths' system is a global optimal symbiotic environment in the vicinity of the Earth's life as a whole.

## § 2.4. DEVELOPMENT OF PRO-MICROBIOTIC ENVIRONMENTS
**Index: Geochemical development of the early biotic Earth, development of surface-dwelling Earth's type of life**

Solar luminosity and atmospheric composition and transparency are continuously changing. Co-developed in changing conditions, present-day biogeochemical cycles and composition of the Earth's compartments have recorded successive differentiation (epigenesis, compare to the egg epigenesis) of the Earth's system. Biogeochemical cycles include the perpetual mineral recycling. Mineral recycling on the Earth's surface relies on the input of energy and matter from under earth. Rocks migrate upward from the mantle. Exposed to low pressure, low temperatures and liquid water on the Earth's surface, they decompose into minerals. For the Earth's type of life, mineral recycling is an important part of the lifespan development over a multiplicity of temporal-spatial scales. Mineral oils, mineral fats, and waxes, water and original organics had formed the first "food" colloids in the upper ocean and, later, the first soil colloids. Due to their high cation exchange capacity, the colloids usually provide surfaces for saponification and for assembly primordial microorganisms and colloids into the biofilms of the surface- dwelling Earth's type of life.

Development of the early biotic Earth is development of its local and global pro-microbiotic environments. Epigenesis of the early Earth had led to the advent of the oxygenic photosynthesis in the surface-dwelling microbiota. About this time, the oxygen sinks beneath the Earth had been mostly filled out. Free oxygen on and near the Earth's surface, became available for consumption by superficial microbiota. Gradually, advent and later formation of the food webs of the oxygen photosynthetic organisms had processed over the superficial environments. The microbial organisms were able to adjust their metabolism to the dry, cold and even dark conditions of the diversity of the local environments on the Earth's surface and to integrate their activities in the hierarchy of the food webs.

## § 2.5. BIOTIC EARTH AND OXYGEN CONTENT OF ATMOSPHERE
**Index: Food webs co-integration in the Earth's type of life, oxygen photosynthetic food webs, solar factor, radiative balance of atmosphere-climate system, pre-oxygenic Earth**

The food webs of the oxygenic photosynthesis have been integrated into the Earth's type of life which includes many different food webs beneath, on and above the Earth's surface. The optimal balance between energy output and energy consumption in any food web depends on the energy consumption and output in the other food webs. Food webs on the Earth's surface are integrated into the surface-dwelling biota and include phototrophs and chemotrophs (included heterotrophs). Though in different proportions, phototrophs and chemotrophs were able to acquire all Earth's surface. The influence of the surface microbiota and its food webs activities on the atmospheric processing and composition is significant even in comparison to the influence of the under earth activities. For example, we see that reduced volatiles from volcanic emissions are effectively neutralized by oxidation reactions; neutralization can be illustrated by example of the separated from liquid water, reservoirs of reduced volatiles (methane clusters, etc).

Atmosphere-climate system of the present Earth reflects on the dominating role of oxygen acceptor in the surface-dwelling biota, and on the such advanced survival skills of the oxygen photosynthetic biota as (i) seasonality of its growth and production, (ii) alternative oxygenic and anoxic photosynthesis, (iii) symbiotic functioning. Performed on the base levels of the food webs, symbiotic functioning permits the optimization of oxygenic and anoxic phototrophic production. For instance, alternative photosynthesis is an ability to switch from the oxygenic to anoxic photosynthesis in some cyanobacterial species exposed to high levels of $H_2S$ ([13]). Optimizing production proceeds in direction from the Earth's life to instances of the life forms;

from the global environment to local environments, from local environments to biofims, from biofilms to active micro organisms.

Solar irradiance puts constraint on the lifespan of phototrophic microbiota and the growth of phototrophic population. During the geological times, the consistent conditions of solar irradiance stimulated the oxygen photosynthetic energy production in the Earth's superficial life. The stable oxygen content of the modern Earth's atmosphere shows that the biological oxygen emissions into atmosphere are closely equated to physiological oxygen consumption in microbiota and its food webs. The associated energy consumption proceeds in a loop, via synthesis and decomposition of matter into molecules and atoms. Produced in metabolic processes, heat and material "waste" are recycled through the food webs and the chain of environmental processes. Biotic-abiotic surface heat fluxes are important for understanding of the Earth's radiative balance and ozone <-> oxygen transformations in atmosphere. Starting the small scale sporadic and recurrent eddies in atmospheric boundary layer and ocean surface mixed layer, surface heat fluxes impact dynamics of the atmosphere and ocean toward redistribution of the chemical and heat "waste" over the Earth's compartments. Redistribution has seasonal character: it reenacts the optimization for the next to come productive periods. Surface heat fluxes are important for understanding of the Earth's radiative balance and ozone <-> oxygen transformations in atmosphere.

## § 2.6. ELABORATION OF OZONE-OXYGEN TRANSFORMATION
**Index: Flexibility of oxygen storage, early biotic Earth, tradeoff of the sensitive terminal oxidants**

Elaboration of ozone-oxygen transformation has existed since the time of the gradual spread of the oxygen photosynthetic microbiota (microbial organisms) and its food webs over the planetary surface. The intrinsic strategy for the energy elaboration is based on the flexibility of oxygen storage in the Earth's superficial subsystems or compartments – ocean, ice/ snow pack, atmosphere. Each subsystem contains a great variety of the active life agents with alike and different metabolic pathways and different sensitivity to oxygen and ozone.

In opposite to the commonly used definition of the pre-oxygenic Earth as the early Earth before the rapid rise of the atmospheric oxygen in serial of the Great Oxygenation Events, we reserve this name for the most recent state of the Earth's development when the oxygen photosynthetic microbiota just begun/or had difficulty to advance over the planetary surface. Whereas, we can state that early biotic Earth had an atmosphere, we cannot state whether the early biotic Earth had reducing, neutral or oxidizing atmosphere, neither describe how the oxygen content of atmosphere had been changing before the wide spread of oxygen photosynthetic microbiota. Since known hypotheses of chronology of the oxygen content for the Earth's atmosphere don't withstand the severe criticism, we avoid in our study to speculate about the geological scale $O_2$ build-up. Consequently, we attempt to reduce the entire problem of the dependence of atmospheric oxygen content on the Earth's biosphere to the stable (from the past to the present) interrelation between biogeochemical cycles of oxygen, halogens and other terminal oxidants for biology. We conclude that elaboration of ozone-oxygen transformation can be reached through the tradeoff of the sensitive terminal oxidants.

Ozone-oxygen transformations impact composition, transparency and circulation of the global atmosphere. After the wide spread of oxygen photosynthetic microbiota and in conditions of increasing solar luminosity, the oxygen photosynthetic microbial organisms and their food webs have joined parties responsible for the regulation of atmosphere-climate system.

Through the atmospheric medium, it became available to control the surface environments where the oxygen photosynthetic microbiota could not be well established because of the lack of the visual light or lack of nutrients. Superficial environments with a steady lack of nutrients for the oxygen photosynthetic food webs reserve a very important case of the co-integration of the food webs in the Earth's type of life. Commonly they comprise the microbiota of beneath and/ or above the Earth's surface. We don't discuss them in our study, but instead we focus on the extreme environments with a recurring lack of the visual light.

The oxygen photosynthetic microbiota had developed the food webs with inclusion of the large life forms. As the geological records show, suffering consequent extinction of upper level species, these food webs were routinely destroyed at the time of the global catastrophes. Then, with several modifications subdued to the new energy budget of the Earth's type of life, they were built over again. In a metaphysical view, the extinct species ceased to be because they could not or stopped to develop, because they could not anymore to fulfill purpose of existence.

## § 2.7. LIGHT- TENTATIVE SUPERFICIAL ENVIRONMENTS

**Index: Interrelation of biogeochemical cycles, tentative extreme environments meet all requirements for the safekeeping of the early pre-oxygenic phenomena, wintertime and springtime polar Arctic**

Biogeochemical cycles of the terminal oxidants laid foundations for the modern oxygen cycling phenomena. High oxygen content of the planetary atmosphere (above 10%) became the critical factor for the spread over of the surface-dwelling aerobic life forms.

At the present state of the Earth's development, biotic pre-oxygenic phenomena can be accessed

   (i)     Through the interrelation of biogeochemical cycles of the terminal oxidants
   (ii)    In the extreme environments

Pro-life conditions in the extreme environments are mostly relevant for microbiotic metabolism. We refer to extreme surface environments at the high risk of irreversible dissipation as to the tentative extreme environments. In tentative superficial environments, emissions related to the local surface-dwelling biota are few and low and have a seasonable character. We speculate that the dissipation of environment becomes irreversible if sufficient active life component is missing. As it happens in wintertime Arctic, life component may be missing just temporarily. Because of sea ice/ snow pack layer isolation between the land and ocean and atmosphere at polar latitudes, the casual biotic-related emissions of Arctic ocean sharply decrease in winter. Due to the metabolic shut down in the oxygen photosynthetic food webs in the dark and cold time of the year, Arctic superficial environments stop to emit biotic-related oxygen species to ABL.

Type of extreme environments that we select to study is a type of extreme environment with low temperatures and small energy flows. According to the laws of thermodynamics, Earth's subsystems at cold temperatures still may serve as possible sources of the low entropy (high quality, structured) energy. In the subsystems with small energy flows, the Earth's type of life takes low entropy energy from the habitable environment until no residue is left. After that, the metabolism and even organic degradation temporarily shut down. Permafrost environment provides good example of described situation.

Local atmospheres of polar regions influence the development of global atmosphere-climate system more than local atmospheres <patterns> out of the polar regions. For instance, atmosphere of Arctic polar regions serves as a biennale and memory of the NH oxygen production: winter low temperatures, winter accumulation of ozone formed in the (previous) sunlit season, increase in ozone life time up to 3-6 months in polar stratosphere, ozone-oxygen conversion at (next) sunlit upon local halogen catalysts emitted into atmosphere from the surface and under earth, specifics of geomagnetosphere etc.

It is believed that at the early Sun (the faint Sun model), biotic Earth had received less solar energy (UV, visual light, IF). The other reason that Earth's surface had been receiving less extraterrestrial energy was opacity of the Earth's atmosphere: The early inside Earth was intensively heated from decay of radioactive elements, so Earth was much more volcanically active. We will never be able to confirm the opacity of the early pre-oxygenic Earth's atmosphere, but it seems logical to suggest that the thin atmospheric boundary layer already was effectively opaque to shield the paleo surface-dwelling microbiota from electromagnetic space radiation. Present Earth's atmosphere is also effectively opaque to the extraterrestrial radiation. Environmental conditions of the past biotic Earth are comparable to the conditions in the extreme environments that we observe today. The (near-) polar axial tilt means that the polar regions constantly receive less solar energy than the regions at the lower latitudes. At both poles, winter weather is being extremely cold and dry and reproduces the very transparent and thin troposphere. During transition and production seasons for the surface-dwelling biota, opacity of polar troposphere is getting higher due to the increased aerosol loading and cloudiness. In winter most polar regions are covered with sea ice/ snow pack. Bordering winter ABL, polar environments are inhabited by superficial microbiota which is active even at low temperatures. Sea ice/ snow pack, the same as open waters, include metabolic by-product impurities. It is reasonable to suggest that past superficial environments were inhabited by the microbiotic carriers of the metabolic pathways similar to the paleo < survived metabolic pathways> marine microbiota featuring present extreme environments ([11]). The survived paleo marine microbiota demonstrates high viability due to its ingenious metabolic schemes.

*Genesis* delivers a poetic description of the early Earth:
(Genesis 1:1-2) In the beginning when God created the heavens and the earth,
the earth was a formless void and darkness covered the face of the deep,
while a spirit (the same Hebrew word has 2 meanings: 1. spirit, and 2.wind) from God swept over the face of the waters. (Genesis 1:3-5) Then God said, "Let there be light";
and there was light. And God saw that the light was good; and God separated the light from the darkness. God called the light Day, and the darkness he called Night.
And there was evening and there was morning, the first day.

Taken out of "the face of the deep", brought above picture surprisingly fits the conditions attributed to the Arctic region of the last 10, 000 - 70,000 yrs. Present extreme environments of the polar Arctic appear especially tentative at the winter sundown and the spring sunrise. Spring sunrise is a time of the restoration of the active life component. Sunrise and the following transition to the Northern Hemisphere's productive season is accompanying by the various natural phenomena, included the Arctic bromine explosion. In [10] we have reviewed springtime tropospheric ozone depletion as a prompt to the productive settings in the Arctic ocean SML (surface mixed layer).

In winter, polar environments feature a small flow of heat and light energy; they also have unlimited water resource with the heat capacity dependable on the state of water. The successful sunrise comeback of the oxygen photosynthetic biota depends on the pro-life conditions in the ABL and surface mixed layer of Arctic ocean. Such as (i) low surface ozone and (ii) intensification of oxygen fluxes and stabilization of oxygen levels in the surface mixed layer (SML) of Arctic ocean, pro-life conditions are violated. Surface ozone mixing ratios reach levels unsafe for the Earth's type of life. At sunrise, bromine compounds are emitted from the polluted sea ice/snow pack surface, and following tropospheric ozone-oxygen transformations alleviate regional ozone-oxygen crisis.

Global correlation between the microbiotic metabolism and atmospheric turnover for the biological oxidants, and ,especially, oxygen, would be best investigated in the light-tentative extreme environments. In light-tentative environments, physical conditions play subdued roles to the sunlit conditions; the local metabolic production of the reactive trace species shows the apparent sensitivity to the changing light energy.

In general, phenomena of the Arctic bromine explosion reflect on the critical activities of the life component on the surface-dwelling Earth's type of life during and after the long blockade of solar light energy.

Tradeoff of the sensitive terminal oxidants is observed in both south and north polar regions. Despite the fact that the North and South Poles geophysics are very different, sources and mechanisms of bromine, iodine and chlorine pollution are different, - surface microbiotic activities and surface halogen emissions from sea ice and snow pack are recorded (at sunrise) for both regions.

## § 3 ASPECTS OF OXYGEN BIOGEOCHEMICAL CYCLE
**Index: Optical transparency, aerosol loading and ozone-oxygen conversion in Earth's atmosphere, surface ozone depletion as a precursor for survival of superficial organic life**

Earth's atmosphere is an example of the globally shared environment with apparent epigenetic oxygen signature. Paramount importance of the oxygen biogeochemical cycle is expressed in the fast turnover of very large oxygen content in the Earth's atmosphere. Fast turnover indicates on the dominating role of oxygen cycle in the energy budget of the present-day biotic Earth.

Oxygenic photosynthesis is possible because Earth's atmosphere is transparent to the visual light. Optical transparency of the present-day atmosphere is effectively modulated at the molecular scale of the chemical reactions and by suspended particulate matter.

Aerosols impose direct stress at the optical transparency and generate perturbations of the major gaseous oxygen-related modulators. Global aerosol Junge layer at about 20 km altitude is formed by the volcanic emissions. Lower aerosol content is formed mainly by the ocean biotic-abiotic emissions. Heterogeneous recycling on aerosols is important for the life times of atmospheric species. During the Earth's development, aerosol radiative forcing of upper atmosphere and lower atmosphere has preceded to the ozone-oxygen radiative forcing. Excluding suspended matter, the major optical transparency modulators are ozone, molecular oxygen and combined oxygen (water and carbon dioxide). Atmospheric chemical response to the solar variability is explicitly expressed in the fluctuations of atomic oxygen and ozone content.

Free and combined oxygen biogeochemical cycling plays a central role in the management of the "waste" and energy flow of the Earth's atmosphere. The global ozone above the tropopause and the hydroxyl radicals below it determine an oxidative (cleansing) capacity of atmosphere. Atmospheric chemistry and biochemistry of ozone and hydroxyl radicals are closely related: both in troposphere and in physiological medium ozone decomposition is resulted in the creation of hydroxyl radicals ([6]).

Oxygenic photosynthesis upon visual light in the surface-dwelling biota became a main developmental issue for the superficial Earth's type of life because of the high efficiency of the oxygen primary reactions. Oxygen is utilized in the physiological processes of oxygen respiration (organisms with respiratory type of metabolism) and aerobic oxygen-dependent growth (organisms with respiratory-fermentative type of metabolism). Intolerant to oxygen, anaerobic organisms use alternative type of metabolism, fermentative type of metabolism, or carry out anaerobic respiration in which a terminal electron acceptor other than oxygen is used. Although they don't contact with oxygen during their life time, they depend on oxygen indirectly, through the connections between food webs of aerobes and anaerobes.

Oxygen primary reactions in atmosphere include ozone-oxygen conversion and formation of hydroxyl radicals out of water and oxygen. Biotic emissions of nitrogen and sulfur species, halogen species and VOCs (such as hydrocarbons) into atmosphere, successfully provide reversibility of oxygen transformations.

Because molecule of ozone disintegrates molecule of transmembrane protein (lipids, fatty acids, a vital component of the cellular walls, ozone concentrations above critical threshold impose life hazard for the exposed Earth's life. Foremost, surface ozone is a threat for the aerobic surface-dwelling life. By emissions of the ozone depletion catalysts, surface dwelling biota regulates the destruction of surface ozone outside of an organism. Important characteristics of the catalysts' emissions are as following: (i) emitted amounts point on the local target, and (ii) even not being large, emitted amounts can be effective. As we observe in case of the Arctic bromine explosion, if catalytic species have a sufficient atmospheric lifetime, they produce extended effect on a global surface ozone.

Ocean bromine concentrating, bromine explosion at sea ice/ snow pack- atmosphere barrier and ABL bromine propagation can be viewed as operation of sub-hierarchy of complex networks. These large natural networks are built, operated and used by the distributed life agents to control metrics of the biotic Earth and, consequently, surface and near surface ozone-oxygen conversion. Control involves the probabilistic knowledge passed (1) from life to the Earth's life agents, (2) in vertical and horizontal succession, from one generation of the Earth's life agents to the others and (3) within the same generation for the large life forms; the latest becomes possible due to their relatively long lifespan in active phase.

## § 3.1. SURFACE AND NEAR SURFACE OZONE-OXYGEN CONVERSION
**Index: Catalytic degradation of surface ozone, errand role of ozone catalysts (e.g. BrO) in the distributed network of the aerobic life agents**

We infer that atmospheric ozone-oxygen conversion and destruction of the surface ozone had to exist prior to advent of the global oxygenic photosynthesis. The inference means that the early atmosphere contained a pool of chemical substances which caused the catalytic ozone depletion. Ozone is not emitted directly into atmosphere, but formed in sunlit atmosphere and degraded, at most, in sunlit atmosphere (Overnight amounts of destroyed or deposited surface ozone have not been investigated globally).

Some of the substances emitted from the surface and beneath Earth are the natural deflectors of ozone. Related to microbiota, and following its life cycle development, metabolic emissions play a profound role in ozone depletion. Some of these metabolic emissions are excreted from the oxygen photosynthetic microbiota, but others are not. It is doubtless that beneath Earth emissions can be responsible for the fine variations of surface ozone, but selective adjustment of metabolic emissions of the World Ocean is possible. Possibilities of such adjustment usually include chemistry of one or a couple of catalysts which are related to the survival of microbiota in the SML of a local estuary.

Ozone decomposition is catalyzed by the biological metabolites and their derivatives. Surface and near-surface degradations of ozone prevent damage to the surface-dwelling Earth's life and return oxygen to biosphere. As known, sunlight condition alone is not sufficient for the transition to the productive season in oxygen photosynthetic microbiota. Happening in sunlit conditions, massive surface ozone-oxygen transformation conveys the "seasonal transition" message to the local surface-dwelling organisms. In case of the Arctic bromine explosion, along their transportation to middle latitudes, bromine-rich air masses (BrO cloud) deliver appropriate "transition message" to the NH's biota. Together with other organizational messages, it unlocks metabolism in aerobic phototrophs and their food webs.

## § 3.2. VOLCANIC ACTIVITIES AND BENEATH EARTH'S EMISSIONS
**Index: Volcanic activities, supervolcanoe system, solar factor, variations of total ozone, connection of beneath Earth's activities and ocean life activities and emissions**

Volcanic activities bring to view the Earth's life deep beneath the planetary surface. So it had been generally assumed that present-day ( last 10,000 – 400,000 years) volcanic and geothermal activities are significantly weaker than in the previous geological periods of the biotic Earth, how this weakening can be explained from the position of the co-integration in Earth's type of life is unknown.

Volcanoes are source of the volcanic degassing, lava flows and hydrothermal eruptions. All volcanoes emit sulfur, chlorine, and ash particulates. Submarine volcanoes directly introduce volatiles to the World Ocean and warm the deep waters. The combined effect of submarine volcanoes on the surface-dwelling life and atmosphere-climate system has not been investigated. Terrestrial volcanoes are better known for the volcanic outgassing high into the stratosphere. In stratosphere, particulate matter has mixed with water vapor to form the haze of the dimming sunlight aerosols. Haze triggers temporarily decrease in global total ozone and initiate global cooling (volcanic winters and, probably, some of the small ice ages). There is no evidence that volcanic activities regulate ratio between free and combined oxygen in present-day atmosphere. The effects of volcanic eruptions on weather patterns and climatic response to the load of stratospheric aerosols are reviewed in [30].

There are volcanoes and there are supervolcanoes; they are connected into the volcano systems. Volcano systems have impact on each other's activity: significant wakes of a volcano initiate activity in the distant seismic/volcanic area. The major difference between volcanoes and supervolcanoes is in the volume of eruptions and in the formative effect on the Earth's surface-dwelling life and surface environments. Volcanoes form mountains and kill surface-dwelling large life forms in a region; supervolcanoes form basins and may change the climate across the planet. Although supervolcanoes erupted last time in the prehistoric era, it is recognized that supervolcanoes' eruptions are threat to survival of the large life forms on a global scale. Hot springs and geysers give evidence for the present-day activity of the supervolcanoe systems.

It is very important to mention that beneath Earth's activities and beneath Earth's emissions into superficial environments and atmosphere correlate with the external energy forcing - solar magnetosphere. The detailed chronological picture of the immediate Solar cycle 24 (since January 2008 ) is presented online at http://www.solarcycle24.com by VE3EN). Perturbations of solar (and Earth's) magnetosphere, similar to those observed in 2009 – 2010 are consistent with the observed earthquakes and volcanic eruptions. Statistics of solar and seismic data display the mutual influence of magnetosphere perturbations and of the beneath Earth's activities.

While perturbations of solar and Earth's magnetospheres influence the beneath Earth's activities, the atmosphere and surface environments are also impacted. The IMF contact of the Sun and Earth's magnetospheres, allows the management of solar energy supply to the planet. Solar energy supply has a complex structure; it is divisible to the slow and the fast solar wind plasma and pure radiation. Non-linear, close looped, dependencies lead to the situation when hysteresis effect for varying solar irradiance and varying magnetosphere field is created.
We suggest that development of the Earth's system exhibits hysteresis in which beneath Earth's activities are advancing to the atmosphere and superficial environments activities. Thus, they have a leading role in the management of thermodynamic non-equilibrium of the Earth's atmosphere. As though, models that don't acknowledge trends in beneath Earth's emissions are useless for the long-term predictions of atmospheric processing.
In spite of the fact that geological patterns of under earth activities and emissions are known, the interpretation of beneath Earth's emissions in context of biotic-abiotic unity of the Earth's operations has not been developed yet.

In first turn, variations of total ozone can be correspond to the variations of solar radiation in optical and near-optical light spectrum. It is optical and near-optical solar radiation that is used in the surface-dwelling Earth's life. Largest columns of total ozone occur in wintertime polar regions. At high altitudes, ozone is depleted by sulfur, chlorine and bromine species of subarctic volcanic origin ([11]). Near and at Arctic surface, ozone is depleted by halogen species.
In polar Arctic, surface emissions have marine biotic origin. In the extreme superficial environments, even if the immediate volcanic outgassing and massive geothermal flows are absent, under earth life and/or its emission activities become an important local life component. Multi-scale, multiple-step symbiotic functioning occurs in spatially extended extreme environments. For instance, Arctic Ocean has a large undersea ridge called the Chukchi Cap. The Chukchi Cap lies in the polar ice cap area. The cap is rich in oil and gas resources. Sequestration of hydrocarbons, beneath ocean Earth's activities and hydrodynamic regime of Arctic ocean waters, including the thermohaline circulation, are linked to each other through the regional energy budget. Survival and productivity of the Arctic surface-dwelling marine biota depend on

(1) The ocean (incl. sea ice/ snow pack) energy budget and composition, while energy budget and composition are tightly connected to the deep ocean-crust processes and to the inner Earth's energy sources
(2) The solar-use atmospheric processing, atmospheric composition and transparency

## § 3.3. OCEAN EMISSIONS AND OCEAN OXYGEN

**Index: Ocean-atmosphere barriers, ocean layering, Arctic ocean example - stratification of Arctic waters, seasonal changes in Arctic dissolved oxygen and oxygen photosynthetic production of Arctic surface mixed layer, sea ice/ snow pack barrier-filter between atmosphere and ocean**

The World Ocean is not only water reservoir, it is a common habitation of under earth and surface- dwelling Earth's biota. Biogeochemical processes that undergo in ocean environments include sedimentogenesis, concentrating, oxidation- reduction. Concentrating processes lead to the capture of chemical species and elements and ease on metabolic activities of seawater living organisms. Some concentrating processes are directly conducting in the physiological medium; for example, concentrations of heavy metals in organisms are several orders of magnitude higher than those in seawater environment. Other concentrating processes are result of environmental processes upon biotic-related emissions. For instance, gaseous concentrating in World Ocean creates dynamic $O_2$-$CO_2$ system. Presented in so called barrier zones (e.g. water-atmosphere, water-coast, water-bottom, river-sea etc), living organisms intensively exchange energy-matter-information. Metabolic activities, carbon and nutrient cycling in the superficial barrier zones are strongly impacted by the availability of dissolved oxygen. Dissolved oxygen (DO) concentrations depend on many sources and sinks, including air-sea exchange, oxygen photosynthetic production of marine life and oxygen downward flux into the interior of the ocean. Most of dissolved oxygen is found in the photic surface mixed layer (SML). SML is characterized by intensive mixing of waters, oxygen content about 3-8 mL/L and low content of phosphates and nitrates. SML is populated by aerobic and anaerobic biota. In absence of the sufficient local biological, oxygen photosynthetic production, sub-surface oxygen concentrations and overall oxygen gradients between ocean surface mixing layer and atmospheric boundary layer depend on many physical-chemical processes and characteristics (e.g. circulation and mixing, oxygen solubility that depends temperature and salinity). Oceanic free oxygen inventories, biogenic primary production and consumption of free oxygen can be estimated from the dissolved oxygen levels.

One can use an example of SML DO Arctic ocean to discuss general variations of dissolved oxygen. Dissolved oxygen and primarily production data is available for many parts of the Arctic shelf ([25]). Data for the mixed layer shows the approaching to saturated and supersaturated levels dissolved oxygen. While in summertime we assume the photosynthetic (plankton) production of the most of dissolved oxygen, the source of dissolved oxygen in springtime remains questionable. For the polar sunrise mixed layer, low-light and low nutrients conditions favor small, microbial life forms like single cells and their populations (algae). In polar environs, algae and algae-like organisms grow both in the open waters and in pores of sea ice/ snow pack.

The continuing increase of amounts of dissolved oxygen stimulate the spring (late March-April) algae blooming on continental shelves and contribute to all Arctic marine food web.

Commonly, the higher productivity of the surface ocean layer is, the closer to the ocean surface the oxygen minimum lies. In literature, the anaerobic conditions are determined by oxygen minimum layer OML as oxygen estimates below 0.2 ml/L. OML is absent for the high- salinity Arctic. The probable reason is (1) the activities of anaerobic                ,
microbial organisms (activities are very high at the brines-sea water boundary) and (2) a well-pronounced stratification of the water column due to the thermocline and pycnocline factors. The Arctic SML layering structure include ([22] and [25]):

    (1) 0-10 m depth - water freshened due to the ice melting
    (2) 20-100 m depth – seasonal pycnocline
    (3) 20-100 m depth - polar water with the seasonal pyrocline above it
    (4) 100-180 m depth main pycnocline
    (5) Returning Atlantic waters

Coastal upwelling of bottom waters create special conditions at the Arctic shoreline. At this point, we assume that since activities of marine biota are weak in early spring, the amounts of seasonal dissolved oxygen in SML depend only on gas exchange, solubility (temperature dependent) upwelling and mixing. In winter and early spring, levels of DO depend on ocean- atmospheric gas exchange through the sea ice/ snow pack barrier. Sea ice/ snow pack is a part of ocean habitat. As a result of bromine explosion phenomena ([10], [11]), bromine-oxygen tradeoff intensifies oxygen fluxes and prevent ozone penetration into sea ice/ snow pack (ocean) environs.

## § 4 METABOLIC ENERGY TRANSFORMATIONS IN THE EARTH'S TYPE OF LIFE AND DEVELOPMENT OF ATMOSPHERE

**Index: Association of life (transcendental) and organic life (physical); physical division: basic energy-matter resources and a storage of terminal oxidants for biology, oxygen electron acceptor scheme and oxygen management in the Earth's atmosphere, place of the large life forms in the hierarchy of the food webs**

Through all its increasing presence on the planet, life regulates the development of the Earth's planetary system. Impact of life on the energy-matter transformations and development of physical reality is possible due to the <temporal> probabilistic nature of material phenomena.

Even though it seems plausible at the moments differing from the moments of origin, life and organic Earth's type of life have a strong connection with each other. Superficial organisms are organic structural instances associated with life. Material activities of living organisms are supported by sensing (perception) and suitable sensation as a method of exegesis of a sensed knowledge.
Parallel destruction and formation of the temporary associations of life and organic matter determine chemical turnover of the organic Earth's type of life. Destruction of the particular structural instances proceeds after the final dissociation (death) of transcendental from a physical-chemical body of the Earth's living species. At atomic and molecular scale, Earth's type of life can be considered as chemical species in physical-chemical domain. Development of the physical-chemical Earth's system is conjoined with changes in the rates of entropy production and entropy lowing. Though particles and elements and their chemical association are subjects to the material patterns <laws>, patterns seem to have little in common with a transcendental life which is distinguished by its specific property of self-reflection. Due to its transcendental nature, phenomenon of life remains a consensual problem of a human civilized life form. Religious practitioners usually regard patterns as created by transcendental life.

Hence, the Earth's system has acquired the substantial amounts of liquid water on the surface and oxygen-rich atmosphere. Enriched by free and combined oxygen, modern Earth's atmosphere denotes the dominating role of oxygen in the metabolic energy transformations in the organic Earth's type of life. Present-day organic Earth's life could be conditionally divided to the organic life above, on and beneath the Earth's surface. Organic biochemistry is carbon-based. In organic life forms, hydrocarbon molecules are the dominant form of energy storage. Reactivity of hydrocarbons depends on inclusion of oxygen and nitrogen. Energy storages are situated in a body of organism and in environment. Survival of organic life and management of hydrocarbon energy storages assume specific properties of environment. For the Earth's superficial environments, it means existence of water resources.

Water-solvent has apparent advantages over many solvents, that are toxic to carbon-based biomolecular structures. Because of their non-polarity, water-solvent doesn't dissolve hydrocarbons (oils and fats). The essential property of heterogeneous water-mixtures in the Earth's conditions is their thermodynamic non equilibrium. Broad empirical data let us to suggest that the present-day Earth's water resources are (re-) created upon metabolic activities of organic life and represent an important component of the waste utilization, or waste control. Even though waste control is warranted <embedded> in the metabolic activities of the distributed life agents and in the entire scope of the Earth's biotic-abiotic operations, it is very little grasped by anthropogenic factor. Water utilization (cleaning) problems (e.g. molecular dynamics of cellulose-water systems, the Gulf oil spill of 2010, etc.) prove the arrogative character of the existed anthropogenic technologies of energy-matter transformations.

Our understanding of the organic Earth's life can benefit from a simplest abstraction of the Earth's life as hierarchy of the food webs < hierarchy of the complex networks>. The microbial organisms sit at the base levels of the food webs. Furnishing knowledge of microbiota is insignificant. Hitherto, the microbiota of above and beneath the Earth's surface is completely unexplored. Although there is much easier to get factual information about its operations, little is known about the surface-dwelling microbiota.

The valid and practically useful points in exploration of the surface-dwelling microbiota and its food webs appear as follows:

(i)     Energy resources
        Coming from space, light undergoes changes while passing the Earth's atmosphere. Spectrum of the photonic energy (light) is changing gradually over the atmospheric depth. Surface-dwelling microbiota is adapted to the light output of atmosphere (and transverse). Directly or by means of the food web, from the space light energy, surface-dwelling biota generates chemical free energy. While oxygenic photochemistry in biota is an engine of the free oxygen biogeochemical

cycling, there are numerous pathways for the formation of the combined oxygen. For instance, the probable original sources of water and other substances of combined oxygen are the biogeochemical mineralization and photosynthesis by Calvin cycle

(ii)      Terminal oxidants
          As known, oxygen is only one of the several possible terminal oxidants for biology ([20] and [11]). We suggest that at the early stages of the Earth's development, the environmental conditions were characterized by (i) temperature and radiation extremes on the planetary surface and (ii) the life-hostile composition of the atmospheric boundary layer. The amounts of moisture, heat and light greatly differed from the present amounts. Amounts of available free oxygen were different. Considering the ozone toxicity, we could suggest the preceding existence of the elaborated mechanisms of ozone-oxygen transformation. The Earth's biota had to use terminal oxidants other than oxygen: halogens Cl, Br, I and light nonmetal elements C, N, P, S. All of these elements were the first introduced into the prototype metabolic schemes. Prototype metabolic schemes include synthesis of the essential amino acids and are absolutely required for survival of the Earth's organisms

(iii)     Large life forms, populations and structured, large-size multi-cellular organisms
          Earth's life carries on the co-integration of resources by means of the biogeochemical cycling of elements through the above, on and beneath the Earth's surface. Making a practice of the oxygenic photosynthesis upon the visual light, surface-dwelling Earth's microbiota became very successful in the co-integration for resources. Surface-dwelling microbiota transformed the global climate-atmosphere system and created the extensive pro-life surface environments. Even the food webs built above the photosynthetic microbiota are referred to the microbiota above and beneath of the Earth's surface, surface-dwelling microbiota is only known microbiota arisen food webs with inclusion of the large life forms which prey on all levels of the all accessible food webs

## § 4.1. COMPETITION FOR PRINCIPLE RESOURCES
**Index: Principle resources, formation and consumption of the principal resources, oxygen and bromine principle resources, DaisyWorld's conceptual selection of the parametrization and habitable constraints of the Earth's system**

In local environment, competing life agents stay in symbiotic connectivity to produce entropy at the lowest possible rate under the given conditions. In truth, we cannot separate from one local environment to another. In a <conjugal> way, local environment is characterized by optimal rates of entropy production.
Competition for the principle resources is intensified by the faster changes in epigenetic signatures. Competition for the principle resources have different outcomes while taking place on the base and upper levels of Earth's life hierarchy.

Base levels of hierarchy correspond to the foundational types of life and to the organic Earth's type of life. Supposing the organic Earth's type of life, in a simplified view, the division of resources for (i) surface dwelling (mainly cold microbiota, phototrops and heterotrophs) and (ii) under earth microbiota (mainly hot, chemosynthetic heterotrophs) and their food webs is as follows:

(i)     The unshared resources are light energy from space and thermal Earth's energy. These two resources power the Earth's biogeochemical cycling. Earth's life is forced to rationalize its activities associated with energy transformations and optimize consumption (and redistribution) of external energy. Some of activities have an extended, global, character. Optimizing redistribution of external solar energy is starting from the UV absorption by stratospheric ozone. Driven by the solar energy, the atmospheric photochemistry and the photosynthetic energy conversion in the superficial Earth's life and are related to each other through the atmospheric ozone layer. The majority of atmospheric ozone content is above the present-day tropopause. While being an indispensable part of oxygen distribution in the Earth's atmosphere, ozone layer protects the surface life from the dangerous UV radiation. Though the early Earth's atmosphere was produced by the underneath emissions, later ocean emissions from the surface of the biotic Earth pushed tropopause, the boundary of the atmospheric ozone layer, up to the higher altitudes. The incoming radiation is consistent with a solar radiation regime, the ozone distribution is dynamically and chemically regulated by the emissions from the surface and underneath Earth and upon general air circulation in such way that it corresponds to the seasonal and interannual variations of incoming UV radiation. If interannual solar luminosity will continue to increase, the ozone layer is expected to continue growing toward higher altitudes. Optimized redistribution of the present-day solar energy corresponds to the thickness of ozone layer of 220-460 DU.

(ii)    While there is a strong competition for the shared resources, the strongest competition is for the principle resources. Principle resources are resources the most vital for surviving. Principal resources may be limited locally. There is an obvious trend to create the accessible reservoirs for the limited resources. The principal shared resources include hydrocarbons and water which are produced and/or renewed by both types of microbiota. Hydrocarbons and water are the main principle resources for the Earth's type of life.

Principle resources are required for the metabolic energy transformation at the base of food webs; when the independent production of principle resources is not sufficient, the tradeoff takes place. Candidates for the principle resources are oxygen, halogens, and ammonia, sulfur and phosphorus compounds etc. Interrelations of the metabolic schemes are expressed in the interplay and meshing within the biogeochemical cycling, and therefore, in the global atmospheric phenomena. In [10] and [11] we have discussed bromine-oxygen interrelation in biogeochemical cycling.

Oxygen is a principle resource for the food webs of the oxygen photosynthetic microbiota. Oxygen is not appeared to be a limited resource for the surface Earth's life: modern atmosphere is a reservoir for free oxygen. However, ozone-oxygen conversion is a necessary prerequisite for the recreation of life-favorable environment on the Earth's surface. Most of required free oxygen is produced in the surface-dwelling Earth's life by the oxygenic photosynthesis. Oxygen is transformed/ consumed in respiration, combustion and other oxidation processes. Human life form extensively consumes oxygen; it is also produces some of required oxygen in industrial ways. Oxygen is and definitely was a principle limited resource for the Earth's underneath-dwelling microbiota. However, we don't posses content information regarding oxygen biogeochemistry that is proceeding beneath Earth. We know that outgassing from the beneath Earth is reducing; the oxygen sinks beneath Earth are believed to be filled up.

Bromine is a principle resource for the Earth's underneath dwelling microbiota: bromine compounds come to the Earth's surface through eruptive and passive outgassing. Bromine outgassing includes outgassing to the atmosphere (mostly, stratosphere) and outgassing to the World Ocean. Outgassing into the ocean provides bromine for the metabolic pathways in the marine microbiota. Ocean bromine concentrating by paleo marine microbiota indicates on significance of bromine and bromine metabolites in the Earth's type of life.
Bromine is a principle limited resource for the Earth's surface dwelling microbiota. At micro scales, it supports metabolism at the very cold temperatures. At atmospheric macro scales, it catalyzes tropospheric ozone-oxygen conversion correlated with solar radiation and ABL dynamics.
Competition for the principle resources of bromine and oxygen involves tradeoff of terminal oxidants through the elaborated light energy applications (such as associated phenomena of the Arctic bromine explosion).

Competition for the light energy exemplifies the resource competition in the Earth's life. As it was demonstrated in different versions of the one-surface model of DaisyWorld (originally conceptualized by J. Lovelock), the common life-favorable conditions include optimal surface temperatures and optimal combined albedo. Albedo is assumed to be geometrical measure of the average reflectivity; combined (statistically processed) albedo is easy to handle. Optimal surface temperature can be understood as DaisyWorld's habitable constraint, it implies liquid water (e.g., constraint on the liquid water). At the present state of the biotic Earth's development, the Earth's hydrological cycle is an instrument of regulation of the space energy consumption. The hydrological cycle and liquid water reservoirs sustain the Earth's type life. Just as the hydrocarbons reservoirs beneath Earth, surface water reservoirs are principal reservoirs shared in the interests of all foundational types of the Earth's organic life.

## § 4.2. PHYSICAL DIVISION: DIVERGENCE AND CO-INTEGRATION IN THE EARTH'S LIFE
**Index: Infinite dichotomy, physical division: multiplicity of distributed life agents, falseness of the anthropocentric views; resource sharing, differentiation and succession in the Earth's life and in the superficial environments**

*Infinite dichotomy in one* has been continuously discoursed throughout the ages of a human history. *Infinite dichotomy in one* appears as the most important case study of natural philosophy.

In a bulk-population, it is often reduced to the self-identification and questioning of faith and doubt. Through math and apart from math, science and art facilitators denote infinite dichotomy in many possible abstractions and images. Formulated in [11, *Principle of multiple unity and interminability of the Earth's biotic-abiotic operations,* is a physical-world expression of a principle of infinite dichotomy in one.

Infinite dichotomy in one, abstraction of the major concept of the physical world regulation, had been already known to the Greek metaphysicists. Ancient Greek philosophers applied concept of dichotomy in physics, math and philosophy studies. Particularly, dichotomy was applied for the classification of matter, of astronomical phenomena and in a field of the early (pre-Aristotle) Greek psychology. By means of infinity dichotomy, one and centralized life predicates the development of the Earth's life as decentralized, robust and exponentially divergent.
While the observed physical reality conforms to infinite dichotomy, and it doesn't conforms to the anthropocentric views. Evidently, centric on a single life form, views contradict to the Earth's biogeochemical records which provide insight for the early biotic Earth's development upon a multiplicity of the life carriers. Anthropocentric views became source of numerous mistakes, e.g. erroneous opposition of human life forms to the other life forms. Among mistakes there are (1) reports on the "unconsciousness", purposelessness of the natural biogeochemical processing and (2) cancellation of the control factor of life and the Earth's life in the earth science modeling. Control factors of probabilistic nature have been replaced by universal physical laws (like conservational laws) and the latter laws have been replaced by their mathematical abstractions in a form of differential math in 3+1 dimensions. Exploited today abstractions are valid for domains much smaller than the domain which the human life form is able to explore. Even the most advanced fundamental theories have their limits. Due to the dogmatization of several fundamental theories through the social institutions and "organized science" in the last several decades, conscious exploration of nature has been slowed. Brought up as natural philosophers, the facilitators of the physically-meaningful differential calculus and differential analysis (G. Leibnitz, I. Newton, J. Fourier, A. Moivre and others), and geometry (P. Laplace,
H. Poincare, H. Helmholtz, D. Hilbert and others) never dreamed of such an outrageous distortion of their accomplishments.

Distributed life agents include earthy organic life agents. The latter can be put into the categories of the cold and hot microbiota. Cold and how microbiota transform the Earth's environments in accordance with fundamental multiplicity and interminability. After the materials from the inside Earth passed through the biotic-abiotic processing, products of this processing have been incorporated into the biofilms on the Earth's surface and have been available for the superficial microbiota.
The biogeochemical development of the Earth's life has led to the emergence of the surface-dwelling Earth's type of life, pro-life Earth's superficial environments and pro-life Earth's atmosphere. Beneath Earth's emissions, subduction and volcanic activities are the channels of the biogeochemical co-integration in the Earth's life. The unity of the Earth's system is distinguished by adequate divergence and co-integration:

The divergence of the Earth's resources between the cold and hot microbiota
Cold microbiota dominates in the Earth's surface environments. Hot microbiota dominates inside the Earth
The co-integration of the food webs of the cold and hot microbiota. Resource sharing

The big part of surface-dwelling biota belongs to the ascending trophic levels of the oxygen photosynthetic food webs and other "cold" food webs. This biota operates in a limited range of surface temperatures, centered on the global mean surface temperature. It is notable that the mean surface temperature for the present Earth is estimated at $15^0$ C. The temperature of $15^0$ C reflects on the "moderate cold" thermoregulation of the surface dwelling microbiota. The smaller part of the global surface-dwelling microbiota is represented by hot microbial organisms (thermophiles). The hot microbiota has conserved along the global geothermal ridge networks. Majority of the global geothermal networks sits at the mid-ocean and oceanic coast ridges. Associated with geothermal activities, thermophile microbial organisms gain (1) nutrients from the volatiles coming from asthenosphere and (2) energy from chemosynthesis. Division and co-integration of the food webs can be illustrated by example of Yellowstone environments. As expected, in Yellowstone supervolcano region, surface-dwelling hot microbial organisms practice geothermal energy conversion. They are also included in the diet of the local large life forms belonging to the food webs of cold microbiota.

## § 4.3. ORGANIC LIFE AND GLOBAL HYDROCARBON ENERGY FLOW
**Index: Separation of hydrocarbon and water reservoirs, recycling of hydrocarbon energy, energy storage out of a body of microbial life forms, on-demand energy transformations**

The co-integration of the categories of the organic Earth's life can be demonstrated through preemption of hydrocarbon energy in a hierarchy of food webs . Network of the food webs is based on the quality and quantity of the hydrocarbon energy flow. The hydrocarbon energy flow proceeds via a variety of earth biotic-abiotic processes. Due to the overlap of processing, e.g. overlap of the ocean and atmosphere processes, overlap of the ocean, atmosphere and under earth processes, modeling of the respective recycling of hydrocarbon energy should be concerned both with the basic properties of the base-levels microbiota and with stability and development of the different Earth's compartments.

Hydrocarbon energy flows rely on separation of hydrocarbon and water reservoirs. Hydrocarbon and water reservoirs are managed by microbiota: in a first place hydrocarbon and water reservoirs are accessible for microbial life forms, in a second place they are accessible for the large life forms.
Deep-going energy flows in the asthenosphere and the lithosphere and immersion of sedimentary basins provide the best conditions for the formation of oil and gas reservoirs. Since water reservoirs are formed on the Earth's surface, the main water and hydrocarbon resources are separated. Separation of resources prevents power generation associated with hydrogen formation in the hydrocarbon-water reactions in supercritical conditions ([31]).
Oxygen photosynthetic food webs spread over the Earth's surface. Mankind is a large life form which belongs to the oxygen photosynthetic food webs. Consumed (recycled) in the physiological activities of a human organism, hydrocarbon energy is eventually supplied from the other superficial life forms (energy-matter). Essential physiological and mental activities of a human organism require solar source of external energy/ or its imitation. Metabolic activities of human civilized life form refer to the energy sources which are not used directly in physiological activities. Human civilization thrives to produce chemical energy from the hydrocarbons <oil and gas> trapped in the asthenosphere, but instead produces a lot of heat waste (quoting Mendeleyev, "to burn oil means to stoke a stove with banknotes"). The industrial oil and gas production involves tradeoff of the shared Earth's resources: surface liquid water is being pumped inside the Earth.

Separation of reservoirs proceeds on and under earth surface. Significantly enhanced by microbiota activities, separation of hydrocarbon and water resources is predetermined by the physical-chemical properties of water and hydrocarbons (aromatic hydrocarbons especially) at the moderate cold temperatures. For instance, hydrocarbonoclastic bacteria consume petroleum oil from the contaminated oceans and marine systems. There is also an evidence of halogenated organic compounds in volcanoes and vents which are regarded to the abiotic and microbiotic chemistry in hydrothermal fluids rich in hydrocarbons ([12], [7]).

Technological solutions for recycling and accumulation of hydrocarbon and water reservoirs have crucial importance for the survival of the human-like life forms and for the development of civilization. There are several ways of accumulation/ formation of under earth hydrocarbon reservoirs. As opposite to biotic hypothesis of oil origin from the oxygen photosynthetic large life forms, D. Mendeleyev had suggested an "abiotic hypothesis" according to which petroleum was accumulated from carbon deposits originating deep in the Earth's mantle. Whilst it is unknown what part in formation and consumption of oil and gas take the chemosynthetic microorganisms of beneath the Earth, it has been accepted that (1) reservoirs are produced in the conditions of the beneath the Earth, (2) such vast volumes of oil and gas could not be produced by the food webs of oxygenic photosynthesis alone. Mendeleyev's hypothesis is supported by the newly discovered fact that oil production can be enhanced via the utilization of microbiota, e.g. butanol production enhancement by C. beijerinckii. In standard earthy conditions, butanol also is immiscible with water at concentrations higher than 7%, which brings about natural separation between water reservoirs and butanol fuel. Butanol produced by microbiota is an attractive energy source for a human civilized life form: distillation of butanol from the water mixture requires several times less energy than to convert crude oil to gasoline.
Butanol production is an illustration of the acidic methanolysis adaptation required for the simultaneous molecular distillation of water and petroleum. In spite of all that, industrial butanol production in estuaries outside of laboratory environment seems to be inappropriate due to the deleterious implications on cell membranes of the marine and terrestrial life forms different from versifications of C. beijerinckii. Metnaholysis is opposite to hydrolysis – the addition of water to the ester link and breaking it; alternation of esterification and metnaholysis should lead to impaired hydrolysis in the non genetically- modified organisms.

Based on the facts given above, we consider "abiotic" processes of beneath Earth as a result of strong microbiotic existence. The metabolic by-products of the beneath (Earth's) life activities are used in the activities of the superficial Earth's life and vice versa. Chemical products such as water and petroleum are used on demand. Reactions of the water and petroleum chain products in physiological medium provides energy storage and on-demand consumption (transformation). Likewise, oil mining provides an example of on-demand usage.

Reactions of water and petroleum outside of physiological medium, both in under earth and surface conditions, lead to production of the "waste" thermal energy. Thermal energy or heat transfer which is generated in reactions of water and hydrocarbons in the Earth's atmosphere, accommodates the biotic enhancement of the energy-matter transformation.

Changes in recycling of the water and hydrocarbons in a body of the Earth's life continuously impact biogeochemical cycles of elements. Though human civilization is extensively processing both types of reservoirs, e.g. quality of water reservoirs, we are far from understanding of the changes- in- reservoirs' impact on the free and combined oxygen biogeochemical cycles. According to the interminable geochemical records free of anthropogenic factor, global reservoirs are changing very slow in cycling manner. As though, the most present-day changes in reservoirs can be seen as reversible. It is believed that current trends for atmospheric oxygen and the World Ocean are as follows: atmospheric content of free oxygen is decreasing, while the sea level of the World Ocean arises, diversity of the large life forms is expanding by account of the human civilized life form only.

Place of any organic life form in the unity of the physical Earth's system is determined by the role of this life form in energy-matter transformations of hydrocarbon and water resources. In terms of water and hydrocarbon recycling on Earth, possible significance (tradeoff resources, etc.) of the mankind is not comprehended yet. In outer space conditions, anthropogenic applications of water and hydrocarbon reservoirs specify advanced conditions for expansion of the Earth's type of life.

## § 4.4. SUPERFICIAL ORGANIC LIFE AND HYDRO RESERVOIRES
**Index: Capacity of the Earth's life, water-solvent biochemistry, thermodynamic non-equilibrium, micro scale waste utilization-control and emergence of the World Ocean and submissive water reservoirs, entropy lowing upon the activities of the Earth's life**

There is no doubt that "cells are more than the information encoded in their genomes; they are part of a highly integrated biological and geochemical system in whose creation and maintenance they have participated" ([33]). Whatever origin of today's Earth's life forms, they share common biochemistry. Common biochemistry and common chemical recognition < sensing> make interactions of the distributed organic life agents. Alone, chemical recognition is not enough for the successful, purposeful functioning of the Earth's life. Purposeful functioning of the Earth's life involves sensing beyond physical division. Fundamental concepts of the organic life, the Earth's life forms and the biotic Earth as a whole imply an important category of capacity, or mightiness:

Transcendental division: Capacity of a life form to change and to adapt to environmental conditions through interactions with other life forms is determined by its qualitative ability to absorb a transcendental component

Physical division: Capacity of a Earth's life form to change and to adapt to environmental conditions through interactions with other life forms is determined by its quantitative ability to absorb and transform external energy (e.g. into a hydrocarbon energy, etc).

In physical division, capacity can be studied through the chemical interactions between the Earth's life agents. At micro scales, chemical interactions of the earthy organisms proceed via

<molecular> water-solvent biochemistry. At macro scales, chemical interactions of organisms surpass <molecular> biochemistry and proceed via the Earth's biogeochemical cycling. Fundamental concepts frame up empirical <material> procedures that determine their applications. In the reduced material domain, definition of the Earth's type of life is biased the Earth's physical-chemical medium. The water-solvent, human-like biochemistry is imminent to all surface-dwelling life species observed till now. Limits of organic life on the Earth's surface concern with water as a biosolvent. Seemingly, limits of organic Earth's life can be derived from the limits of microbial organisms in solid and liquid phases of water.

Because external energy must be obtained outside of organic body of the life forms, metabolic activities are possible only in environments at thermodynamic non-equilibrium.
A water molecule and association of water molecules carry both nucleophilic and electrophylic centers. It means that in a water medium, e.g. ionic solutes, thermodynamic non-equilibrium is easy- to-create. Water reservoirs are populated by organic life agents and dissolved organic matter. Metabolic processes induce thermal gradients that cause to the reorientation of the water molecules and large polarization fields, and consequently regulate physical-chemical parameters of water medium and ionic solutes' concentrations.

Emergence and maintenance of the World Ocean are uphold by the organic Earth's life because hydrophobic <fatty> membranes allow to accumulate water outside of a body of organism. We suggest schematic interpretation of relation between the global scale hydro cycle and waste control at the micro scale of a living cell. Accumulation of water outside of a body of organism is envisaged as following: (1) part of water content is chemically produced in cells, (2) metabolic waste in a solute form (and occasional matter that has been diffused into the cell) is removed from organism by effix mechanisms. (3) As decreased water content of the cells puts limits on organic life, metabolic activities (1) start over again.

In absence of the major beneath Earth's and outer space events, through its carrying energy capacity and ocean emissions, the World Ocean life medium determines the pro-superficial-life properties of the immediate atmosphere-climate system (and especially ABL, troposphere and lower stratosphere). Via complex processing of energy exchange, atmospheric regulation of thermo and radiative forcing at the Earth's surface-dwelling life proceeds toward entropy lowing. Energy and entropy (concept of statistical energy) considerations are central for the physical earth and life sciences. Entropy lowing becomes available at thermodynamic non-equilibrium. Inasmuch as biotic Earth and its subsystems are natural dynamical systems, they have factors with arbitrary small entropy. As though, they can resist to the entropy growth. Since all physiological activities of the Earth's type of life represent extraction and transformation of low entropy from the environment, long-lasting entropy lowing makes developmental progress more efficient. While increase of entropy of a physical system is spontaneous, the reduction of entropy, or structuralization of environment, takes an effort. Provided by the Earth's life, the elaboration of energy transformations is a leading component of the enforced entropy lowing in the Earth's subsystems and in a body of the Earth's life. In general, entropy-like quantity can be drawn from the energy-information equivalence both for a particular energy transformation and for the entire natural phenomena. Specifically, entropy function, rates of entropy production and loss can be estimated through the observed physical parameters of ozone-oxygen transformation upon Arctic bromine emissions (see the future parts of this study).

# § 4.5. MICROBIOTIC METABOLISM AND PRE- AND OXYGENIC PHENOMENA

**Index: Dependence of the surface-dwelling biota on prototype metabolic schemes, bromine as terminal oxidant for biology, bromine metabolites and their derivatives, ozone recycling and emissions of terminal oxidants (bromine) into oxygenic atmosphere**

Not by chance, the maintenance and recreation of the oxygen-rich Earth's atmosphere are related to the metabolism of paleo marine microbiota in the extreme environments:

Many biological species do not produce the sufficient amounts of essential amino acids independently ([11]). Through the multiple unity of biotic-abiotic operations of the Earth's system, and specifically through the symbiotic by their nature food webs, modern biota shows its complete dependency on the paleo metabolic pathways evident in activities of microbial organisms. For example, it was found that methyl bromide (MeBr) uptake in soils must be considered completely microbial ([9]); e.g. at very high amounts of MeBr ( >1 ppmv), methyl bromide is intensively transformed by methanotrophic and nitrifying bacteria over the diverse consumption activities. There is an indication that the metabolic utilization of methyl bromide is a constitutive process - for typical 10 pptv of MeBr, in the unfumigated soils methyl bromide is completely consumed from the soil surface by bacteria metabolizing aerobically (oxygen tolerant bacteria).

> The symbiosis of the Earth's biological species is governed by the prototype metabolic schemes. Before the widespread of oxygenic photosynthesis became available, the Earth's underground, ice and deep water microorganisms utilized the heat energy and the EM energy. Metabolic utilization of terminal oxidants other than oxygen was essential for survival. Many elaborations of energy-matter transformation can substitute each other to a certain degree, however increase of the number of choices to survive guarantees the interminable development of the Earth's life. The pre-oxygenic phenomena of the biotic Earth are preserved in the superficial extreme environments for the future developments

Superficial environments of the biotic Earth simultaneously consume and produce gaseous chemical species. They can consume and produce much faster than gas is exchanged with the atmosphere. Although the net flux, e.g. oxygen flux, could be small, it often results from the near- balance of two opposite-directed gross fluxes over the atmosphere and the Earth's surface (open ocean, ocean covered by sea ice/snow pack etc).

In case of oxygen, dynamical gross of distributed fluxes is really large, and the oxygen content and layering of the present-day Earth's atmosphere may not be as stable as assumed in first approximation. Nevertheless, free atmospheric oxygen consists of molecular oxygen and odd oxygen where the latter is presented by atomic oxygen and ozone; atmospheric circulation and *odd oxygen <-> molecular oxygen* transformations provide the life-sustainable oxygen concentrations in the global atmospheric boundary layer. In the upper atmosphere, atmospheric circulation and ozone-oxygen transformations make up the ozone layer UV radiation shield.

At the present state of the Earth's development, ozone recycling is a most important part of the oxygen biogeochemical cycling. Biogeochemical cycling proceeds via multiple-scale periodical phenomena. Atmospheric ozone recycling heavily depends on the solar dynamo factor and perturbations of IMF. It directly corresponds to the periodicity of the solar dynamo.

Owing to the synchronization with external (e.g. solar) energy flows through the Earth's microbiotic habitats in hydrosphere, atmosphere, soils etc, Earth's type of life successfully manages the Earth's atmosphere-climate system at global and regional scales. Synchronization of biotic-abiotic processes determines the temporal development of the present-day Earth's surface. Concerted phenomena can be approached at diurnal, seasonal, annual, etc. scales. Because of the distributive processing, long-lasting phenomena always would have a global influence. For, in instance, in [11] we have discussed interactions and the synchronization of the biogeochemical cycles of oxygen and bromine species taking place in the immediate marine boundary layer (MBL) of the Arctic troposphere at the polar sunrise. In case of the Arctic bromine emissions, we speak about seasonal and interannual periodicity of phenomena and its significance for the Northern Hemisphere's oxygen photosynthetic production of hydrocarbons in the surface-dwelling microbiota and its food webs.

## § 4.6. GAIA'S CAPACITY FOR CHANGE
**Index: Physical division of Gaia: capacity for change, possible quantitative changes over the biogeochemical cycling**

The great portion of recycling happens on the safe distance from the superficial environments dominated by the oxygen photosynthetic food webs -   above in upper atmosphere, deep in ocean, at the remote polar superficial locations. Through the volcanic eruptions it is regulated by the beneath Earth microbiota. We may guess that it is also regulated by the above Earth, air-born stratospheric Earth's biota, yet absolutely unexplored. The ozone recycling illustrates co-integration in the Earth's type of life, or the fundamental law of interminable multiple unity of Earth's biotic-abiotic operations.

It appears that Earth's type of life has almost unlimited capacity for change, possibility for the almost unlimited increase of complexity. It is reasonable to conclude that increase of complexity is unachievable without (i) a purposeful control factor and without (ii) a good system design. Increase of complexity (iii) requires the growth of energy consumption and efficient entropy lowing, e.g. introduction of the elaborated energy technologies.

We opine that oxygen biogeochemical processes expanded over the Earth's surface later than the other major elemental biogeochemical cycles: the dramatic emergence of the present biogeochemical cycling and oxygenic atmosphere became viable after oxygen sinks beneath the surface filled out essentially. Oxygen biogeochemical cycle is a part of the Earth's biogeochemical cycling: it is tighten with other biogeochemical cycles, especially with cycles of terminal oxidants for biology (carbon, etc.) For example, carbon geological sequestration beneath the Earth's surface is continuing (e.g. due to tectonic drifts), which means that carbon and, consequently, oxygen atmospheric contents are vulnerable to the geological development of the inner Earth*.

In conditions of increasing solar luminosity, tight coupling of the elemental biogeochemical cycles has led to abundance of oxygen at the Earth's surface. As things are, the main reservoir of combined oxygen is the hydrosphere (as 1.2 e12 Tg) and the main reservoir of molecular oxygen is the Earth's atmosphere ocean (as 1.2 e9 Tg). Although oxygen and ozone are known as trapped species in ice, the free oxygen content of the Earth's sea ice/ snow pack has not been separately estimated. The photolytic production of $O_2$ in ice occurs due to incident UV photons, low energy electrons and energetic ions. Oxygen photolytic flux from the Earth's sea ice/snow pack has not been investigated properly. One can guess that photolytic production may be considerable source of oxygen in the clouds of upper atmosphere. However, for the winter and springtime polar Arctic, sea ice/snow pack oxygen flux into ABL is negligible compared to the oxygen transported with air masses.

*Henceforth, it mistaken to make a direct link between the anthropogenic carbon emissions to the global climate changes. The more likely, in yet unknown conditions, the anthropogenic activities such as the oil and gas mining, or beneath Earth's explosions may become a trigger to the global climate changes

## § 4.7. ATMOSPHERE AS AN INDICATOR OF METABOLIC OZONE- OXYGEN SENSIBILITY IN THE EARTH'S TYPE OF LIFE

**Index: Concentrations of ozone and oxygen in troposphere, variations of ABL oxygen, global ozone-oxygen management in interests of the surface-dwelling life, impact of reactivity of free oxygen (oxygen and ozone) on cellular activities and severance of micro climate layer in vicinity of superficial biofilms**

As easily seen, biotic Earth has attained oxygen global management in interests of the surface-dwelling life. In fact, present-day biotic Earth controls temperature profiles of the Earth's atmosphere-climate by means of re-creation and alternation of the global atmospheric reservoir of oxygen and oxygen biogeochemical cycling.
The surface of the pre-oxygenic biotic Earth was gradually conquered by the oxygen photosynthetic microbiota. As a result of the free oxygen accumulation in the atmosphere, pre-oxygenic biotic Earth received a chance for development of the large surface-dwelling life forms, comprising the final development of a human civilized life form.
Sensibility to oxygen is everything that makes sense about the oxygen photosynthetic superficial biota. Before the total area of extreme environments on the Earth's outer surface had shrunk to present-day area, the emergent metabolic schemes in superficial microbiota of the extreme environments had provided the Earth's transition to the modern biogeochemical cycling. Now, following the annual solar cycle, the same metabolic schemes provide the Earth's transitions from one season to another.

Atmospheric processes indicate on the metabolic ozone-oxygen sensibility in the Earth's type of life. Modern oxygen cycling is characterized by (i) the high oxygen content of ABL and (ii) ozone accumulated above tropopause.
Tropopause is assessed from the atmospheric temperature and moisture profiles. Climatological data, tropopause height and tropopause mixing processes are described in [5]. Present-day tropopause is located at altitudes of 8-30 km. The influx of short-wave radiation into the upper stratosphere initiates photochemical ozone production. Because, determining by the reaction rate, the lifetime of ozone is short compared to the transport times, ozone accumulation takes place. Ozone is a radiation-modifying component of the Earth's atmosphere. In parallel, it affects the distribution of atmospheric reactive oxygen species like $NOx, SO_2, CO_2$ and the radicals, including the universal atmospheric cleanser hydroxyl radical OH.

We try to project contemporary settings of oxygen cycling backward to the early biotic Earth. Inside of the oxygen biogeochemical cycling ozone could be considered as an oxygen source and vice versa. It is likely hopeless ever to determine "which came first" the oxygen or the ozone. The possible origin of the early Earth's atmospheric oxygen is water vapor and other complex chemical compounds emitted from the Earth's surface and beneath and then underwent photochemistry and lightning. The possible origin of the early ozone is the same as contemporary origin of ozone in the free atmosphere – through the photodissociation and reactions with NO and radicals. Reversible ozone-oxygen conversion for the free atmosphere and the atmospheric boundary layer had been proceeded in unknown atmospheric conditions. Whatsoever, as a result of this reversible conversion, early Earth had developed into the biotic Earth with its pro-oxygenic-life superficial environments. Pro-life superficial environments had received enough of light energy in optical and near optical diapasons. In relation to the global physical-chemical situation, the oxygen photosynthesis was the most elaborated metabolic application of light energy, and pro-life superficial environments became inhabited with the food webs of oxygen photosynthetic microbiota.

There are a few differences between organic Earth's life species in their sensitivity to ozone. The hypersensitivity to ozone is explained by the organic reactivity of ozone. Formed from ozone and organic molecules, secondary oxidants have ability to destroy the protein component of the cell membrane. Ozone also reacts with reminiscences of organic matter – it scavenges sulfates, denitrificates urea, oxidizes organic carbon. Pre-oxygenic biotic Earth had to have the protective atmospheric layers, in which catalytic ozone depletion and ozone-oxygen transformation took place.
Being the only atmospheric layer directly determined by the metabolism in the superficial oxygenic microbiota, the early atmospheric boundary layer would have been a chief location of the ozone depletion.
Metabolic activities of microbiota control diversity of the micro climate layers and micro environments. Due to the divergence of the life forms development of the food webs, micro climate layers and environments significantly overlap and extend. At present, large populations and large life forms of oxygen photosynthetic food webs easily extend their specific (oxygen-rich etc.) micro climate layers over the global atmospheric boundary layer.

## § 4.8. ATMOSPHERIC PROCESSING VS. METABOLIC FUNCTIONING
**Index: Physical division: Parallel relationship of atmospheric and metabolic processing, functioning of the Earth's-scale unity of the life forms**

Analysis of a parallel relationship of atmospheric and metabolic processing allows to reconstruct correspondence of biogeochemical cycling over the range of the Earth's scales**.** Parallel relationship includes but not bounded to the following functional phenomena required for survival and development of the Earth's type of life:

1) Actualization: Molecular scales: Oxygen and derivative oxidants in oxygen-nitrogen mixture decidedly catalyze metabolic chemistry in the surface-dwelling biota. Global scales: Metabolic chemistry in the surface-dwelling biota intensifies hydrocarbon energy flow through the biotic Earth. Terminal oxidants for biology and their derivates have a fast turnover in superficial environments and atmosphere; biogeochemical cycling of the terminal oxidants has the strong influence on the global energy budget and entropy lowing in the Earth's system

2) Structuralization, establishment of certain topology(-ies): Molecular and global scales: Physical-dynamical profiles of superficial environments are formed due to the specific atmospheric layering, composition, transmittance etc. Earth's atmosphere is a structured environmental compartment, which processing is resolved by metabolic emissions in response to the external energy forcing. Metabolic by-products undergo transformation and storage in the Earth's interior, the Earth's atmosphere and superficial compartments adjacent to atmosphere. As observed, waste recycling of the fast-turnover metabolites / metabolic by-products which are required for the metabolism of oxygen photosynthetic microbiota and its food webs is almost solemnly performed over the global, multiple-layers structure of the Earth's atmosphere

3) Self-regulation: Molecular scales: Periodic input/ output of trace volatiles (e.g. bromine) regulate the cell cycle. Global scales: Diurnal and seasonal influx/ outflow of trace volatiles regulate the atmospheric processing. Related to the metabolism in surface-dwelling microbiota, the net input of volatiles was smaller for the early biotic Earth than at present. If so, the atmospheric boundary layer which we associate with today's pro-life conditions of surface-dwelling organisms was thinner, and tropopause was lower. We can't say what was the oxygen content of atmosphere, but we understand that for any oxygen content, the open lower tropopause means the increasing ozone levels in the atmospheric boundary layer below it.
Aware of a verity that above the critical threshold, ozone levels are dangerous for the organic Earth's life, one would expect particular adaptations in the paleo microbiota. Aimed on the surface ozone depletion, metabolic emissions are an effective adaptation in paleo marine microbiota. In present polar Arctic, during the bromine explosion season, we are able to observe this particular adaptation in action ([10],[11]).

4) Communication: Molecular scales: Signaling molecules and their receptors serve for the inside and outside of cell communication. Global scales: Atmosphere as a communication medium serves for the physical-chemical information exchange between the life agents distributed over the superficial environments

5) Defenses: Molecular scales: Defensive proteins, immune system etc. Lipids (fats etc) make up cellular membrane that constitutes a barrier for chemical compounds placed in and out of cell. In multi cellular organisms, lipids serve to store energy and mediate communication between cells. Global scales: Differentiation and succession in the Earth's type of life and Earth's environments, elimination of toxic concentrations of chemical species: waste recycling, chemical species are transformed throughout biotic-related chemical and photochemical processing. Micro climate layers and ABL ( water content, carbon oxides and carbohydrates, etc.) constitute a barrier for chemical compounds in and out of a body of the superficial biota, serve to store energy and mediate physical-chemical communication between the distributed Earth's life agents

Apparently, compared metabolic and atmospheric phenomena tell us about the united nature of the Earth's biotic-abiotic operations. As much as we are concerned, from own perspective on the interoperability, there is no principal difference between functioning of the Earth's type organisms (e.g. human-like organisms) and functioning of the Earth's-scale unity of the life forms.

## § 4.9. CONDITIONS FOR THE SURFACE OZONE ELEVATIONS
**Index: Elevated levels of surface ozone in polar Arctic environment**

Suggesting the arise of the first oxygen photosynthetic environments in water medium and global coverage of the Earth's surface by the World Ocean, we must assume sufficient amounts of dissolved oxygen in photic zone of water column. In the coldest and darkest regions on the planetary surface, the oxygen fluxes into the World Ocean become available (i) due to the oxygen emissions from/ through the sea ice/ snow pack and/or (ii) due to the oxygen flow from the atmospheric boundary layer.
Elevated ozone levels in the absence of photochemistry is one of the major survival problems for the early biotic Earth (and especially at the times of supervolcanoe eruptions).

Underwater eruptions reshape Arctic ocean floor, ocean waters absorb volatiles and heat of inner Earth. Some volcanoes (Alaska, subpolar Iceland) erupt into sea ice/ snow pack and high atmosphere. Now and then, many eruptions are sulfur-rich; they are resulted in injection of vast sulfur amounts to the stratosphere. Influx of sulfur species is efficient in activation of halogen (chlorine, bromine, iodine) chemistry and, consequently, destruction of stratospheric ozone**.** However, eruption has a number of side effects working to increase ozone: for instance, having volcanic lightning is usual. Lightning could have given rise to the surface ozone above the threshold**.** Though causal connection is unknown, it was noticed that Iceland volcanoes erupt at the same time that northern lights are visible < it is not clear what is a cause and what is a consequence in a causal connection of lightning, polar auroras (electromagnetic plasma outbreaks) and associated phenomena of volcanic eruptions >.

During the polar winter, at minimum of the visual light energy, Arctic ABL is a kind of open-top chamber: the injection of the stratospheric ozone lead to the high variability of the surface ozone and increase of ozone concentrations in hundreds percents. Polar Arctic sunrise features low and high frequencies of photochemical episodes. Elevated surface ozone levels are formed by the ozone down lift and absence of photochemistry.
At low latitudes, high levels of surface ozone happen because of sunlight, warm temperatures, winds and presence of ozone precursors of local origin such as nitrogen oxides. For dark and cold season at high latitudes, in absence of the sunlight, high levels of surface ozone happen because of stratospheric injections and ozone accumulation. Unfortunately, there is a significant difficulty to determine ozone precursors of local Arctic origin. Since then, it has been believed that Arctic surface ozone precursors are related to the anthropogenic pollutants that reach from the outside of Arctic and to the local ship and low height aircraft pollution. The review of surface ozone data and some climatology is given in [35], [3] and [14].
One can easily see, that depletion of the surface ozone at polar sunrise and depletion of polar stratospheric ozone are principal channels for the oxygen biogeochemical (re-) cycling in the Earth's atmosphere, and as though phenomena must have existed prior to the emergence of the undeliberate anthropogenic emissions.

## § 4.10. SURFACE OZONE AND HALOGEN-CONTAINED BIOFILMS

**Index: Properties of halogen compounds in diverse media, atmospheric vicinity of halogen-contained biofilm, local ( e.g. Arctic) ozone-oxygen management in interests of the surface-dwelling life**

Halogens are trace elements in the biogeochemical cycling; except fluorine, they are also terminal oxidants for biology. Though settings for the trace elements cycling can't be projected backward into the geochemical history, we can learn about these setting from the present-day data provided by microbiology and earth sciences (atmosphere, ocean, soil, etc.)

The detailed information of surface ozone field and detailed phenomenology of Arctic bromine explosion is brought in [10] and [11]. We have assumed the significance of the halogen metabolites, and have regarded them as precursors for ABL ozone-oxygen transformation. Further on, e. g. in modeling, Arctic bromine explosion is treated as an expression of a major metabolic adaptation.

During the productive season, flux of the biotic halogen metabolites from the open ocean is generally presented by fluxes of monohalomethanes and their derivatives. The situation with fluxes is more complicated outside of productive season. Biotic halogen emissions from the Arctic sea ice/ snow pack surface undergo the intermediate processing - heterogeneous sea ice/snow pack chemistry and photochemistry. The importance of the intermediate processing is in its temporal coincidence with surface ozone concentrations above the critical threshold for the oxygen photosynthetic microbiota. At polar sunrise, halogen (bromine, chlorine, iodine) influx to Arctic ABL cause to the complete depletion of surface ozone. Ground-based influx of bromine compounds into ABL correlates with downward intrusions of stratospheric ozone when polar vortex is disrupted.

At sunrise, the depletion of surface ozone is determined by emissions of bromine metabolites from the sea ice/snow pack surface. Emissions of bromine metabolites can be viewed as a response of the natural complex network of the distributed life agents to the ozone-oxygen stress. Empirical expression for the surface bromine flux has presented this dependence in the transverse form [10].

It is quite natural to look at the bromine-polluted Arctic sea ice/ snow pack as at a biofilm extrapolation. For the biotic-abiotic multiple unity of the Earth's operations, halogen-contained biofilms are especially important for the very cold icy surfaces of interactions:

(i)     In atmosphere, bromine compounds are efficient at destroying ozone. High levels of the surface ozone are detrimental for the food webs of the surface-dwelling oxygen photosynthetic microbiota. At polar sunrise, Arctic bromine explosion leads to the complete depletion of the surface ozone and the intensification of oxygen fluxes. Depletion of surface ozone and intensification of oxygen fluxes are required for the optimal environs. Based in the GOME data and model partitioning of $Br_y$ ([10]), the total mass of Arctic ABL reactive bromine is estimated around 2 tonnes ( model chemistry includes recycling, but doesn't include production on aerosols)

(ii)     At low temperatures, bromide (iodide) ions catalyze ester saponification reactions (so called halide exchange mechanism)

(iii)    Halogen-contained biofilms could be responsible for the liquid water retrieval. It was I. Langmuir who discovered that the introduction of particles of dry ice and halogen species (iodide) into a cloud at low temperatures induced a chain reaction led to raining or snowing. Hypothetically, halogen-contained biofilms may exist on the stratospheric clouds. Up to now, only abiotic bromine content of stratospheric clouds is confirmed. In August 2008, the Kasatochi Volcano (Alaska's Aleutian Islands) injected volcanic BrO directly to altitudes of 8 - 12 km. The total mass of reactive bromine released into the troposphere/lower stratosphere was estimated around 50 - 120 tonnes, which corresponds to approximately 25% of the previously estimated total annual mass of reactive bromine emitted by volcanic activity (GOME-2, European Space Agency (ESA)). From here it follows that mainly, stratospheric bromine content has the beneath Earth's origin

(iv)    Bromine effect on thermoregulation, cold acclimation and freezing stress in living organisms is explained (a) by the good solubility in water for bromine and bromide salts and (b) by low freezing point (for Br2 -7.$^{20}$ C, for BrCl -6$^{60}$ C) – addition bromide salts to water make the freezing point of water solution significantly lower. Bromide metabolism pathways also induct the enhanced tolerance to the cold in the large organisms like fish

## § 5 OXYGEN EPIGENETIC SIGNATURE OF A MODERN ATMOSPHERE
**Index: Atmospheric metrics**

A common interpretation of the control factors(s) for the Earth's system can be given in terms of the measurable characteristics of the multiplicity of the Earth's life and Earth's environments. Estimates for a statistical control factor of the surface-dwelling Earth's type of life are the estimators biased (1) in the sensitivity to the physical conditions and to the chemicals for the base unicellular levels of the Earth's type of life and (2) in the presently observed state of the Earth's atmosphere <which state is a counterpart of this sensitivity>.

Atmospheric metrics are the physical-chemical and topological metrics of global atmosphere. Like global metrics corresponds to the sensitive dependence of the superficial Earth's life as a whole, sensitivity of the local Earth's life component corresponds to physical-chemical metrics of the local environment. Distributed life agents communicate atmospheric metrics in the hierarchy of complex networks. Any "message" and any link-state (topology) created by the distributed life agents in the atmospheric domain utilize the atmospheric metrics to evoke the best possible metabolic pathways (enzyme concentrations, emissions etc.) and to rebuke symbiotic connectivity. Some of the atmospheric metrics, namely epigenetic signatures, have more significant impact on the development of the biotic Earth and processing of the global energy-matter budget. Reasonably, because epigenetic signatures changes very slow, changes in energy-matter budget, and particularly global climate change can be always traced back by help of the geological records and paleontological data on epigenetic signatures.

## § 5.1. EPIGENETIC SIGNATURES AND MULTIPLE-ELEMENT LIMITATION IN THE EARTH'S COMPARTMENTS

**Index: Multiple-element limitation hypothesis**

"Living organisms are the main factor determining the migration of chemical elements on the earth", V. Vernadsky in "Biosphere and Essays on Geochemistry"

The same as the Earth's atmosphere has an epigenetic oxygen signature, the World Ocean has its own epigenetic signatures. To suggest possible epigenetic signatures, we have extrapolated ( TRIZ, Altshuller) the multiple-element limitation approach (MEL) to the plant nutrition ([27], [28] and [29]) over the Earth's life on the global scales. Extended MEL hypothesis puts "elemental" ratio correspondence between (1) critical levels of chemical resources in body of the Earth's life and (2) observed levels of chemical resources in environs. High elemental ratios of sulfur S, phosphorus P and halogens Cl, Br, I in marine microbiota call forth to propose that epigenetic signatures for the World Ocean might be associated with sulfur, phosphorus and halogens. Critical levels are determined from cellular physiology, and thus, primarily relevant for the diversity of unicellular microbiota.

> In case that base levels of the Earth's life's hierarchy remain intact, the competitive replacement of some marine life forms by the others with more favorable distribution of active life component does not really change the critical levels of chemical resources in a body of the marine life.

Present-day epigenetic signatures for the global Earth's compartments give a complete statistical description of the Earth life and its categories. Set of the ocean epigenetic signatures describes ocean as a pro-life environment for both the under earth and surface-dwelling Earth's life. Set of the atmospheric epigenetic signatures describes atmosphere as a pro-life environment for the surface-dwelling, under earth and above earth microbiota and its food webs. Epigenetic signatures indicate on:

(1) Life and the Earth's life's abilities of elaboration of the energy-matter-transformations and following redistribution of internal resources
(2) Compensatory (tradeoff) redistribution of terminal oxidants for organic biology
(3) Competitive differentiation and co-integration between the foundational categories of the Earth's type of life toward an acquisition of the principle resources
(4) Purpose of existence of the surface-dwelling life agents at the present state of development is an acquisition of the external solar radiation in the superficial habitations

Taking together, epigenetic signatures of environment and elemental ratios in a body of the Earth's life forms constitute metrics of the Earth's type of life. At the moment, we know only several epigenetic signatures of the tropospheric global environment (e.g. oxygen-ozone). From the perspective of the Earth's system's development, tropospheric and surface ocean epigenetic signatures put strict limitation on the survival and development of the large life forms that inhabit Earth's surface. Because critical levels of chemical resources differ from the one life form to another and because large life forms (e.g. human-like life forms) can exist only in the system populated by microbiota, selection of a life form that would be appropriate for the Earth's development, is made at the level of the quorum sensing of the Earth's life as whole, and then a selection is communicated through the physical and transcendental divisions.

In agreement with the MEL approach, local elemental shortages of resources have a potential to initiate changes over the global epigenetic signatures. Microbial life forms that are capable to regulate (keep or modify) epigenetic signatures for the liable supply of external energy, contribute to the sustainability of critical resources and the sustainability of the Earth's life as a whole ([1]). It is important to mention that there is a broaden deviation in the optimal elemental ratio within biofilm and cells (e.g. a liter of human blood can dissolve 200 cc of oxygen gas, which is much more than water can dissolve - oxygen saturation between 65% to 99%). This deviation means that concentrating (sequestration etc.) in environs can go very different throughout the biogeochemical development of the Earth's system and even throughout seasonal biogeochemical cycling. Since there is strong life forcing vigorously favoring life forms able to facilitate elaboration of energy transformations, change of the epigenetic signatures is a key challenge for a life form. Processing of the epigenetic signatures in the tentative environments is essential for the gradual change of the epigenetic signatures. It is a perspective that we glance at the critical processing of oxygen and ozone content in the sunrise Arctic ABL.

Energy exchange and entropy lowing in the Earth's system and subsystems progress at abundance of chemical matter and external energy. Even though, superficial Earth's life as a whole is not experiencing any significant shortage of chemical matter, in some critical environments on the Earth's surface distributed life agents encounter a shortage of water-solvent and other local shortages. In Arctic, distributed life agents encounter a shortage of the external energy.

## § 5.2. CRITICAL PROCESSING OF ARCTIC OXYGEN AT SUNRISE
**Index: Oxygen metric, observed atmospheric concentrations and critical thresholds in the Earth's life for the reactive bromine and oxygen species, homeopathic (small dosage) considerations for the metabolic tradeoffs**

We suggest that paleo metabolic pathways are preserved in a body of the Earth's life. If necessary for interminability of the Earth's life, topological databases of complex networks for the < latent> metabolic pathways are reactivated. Topological physical-chemical databases also support development of the Earth's system because they allow comparison between the dormant and the newly active pathways. It is a quite apparent that sensitivity of the Earth's life accounts for the topological databases corresponding to the global metrics of the past states of the biogeochemical Earth's development.
Earth sciences and (micro-)biology data entails an important contribution of halogen chemistry to the establishment and progress of the global ozone-oxygen management. Associated phenomena of Arctic bromine explosion is processed through one such topology, periodically reactivated by solar factor. This topology is very specific because it is straightforward corresponds to the microbiotic activities <the base levels of hierarchy of the Earth's life>.

For instance, bromine and oxygen biotic-related emissions influence Earth's atmosphere-climate system via range of the chemical and physical processes. Transport-relevant emissions, they influence atmospheric environment depending on where and when the emissions are introduced into it. Bromine and oxygen are terminal oxidants for biology; bromine and oxygen chemical compounds belong to the principle resources shared by the foundational categories of Earth's type of life. Atmospheric amounts of oxygen/ ozone and bromine reactive species befit for the sensitivity in the superficial organisms, comprising oxygen photosynthetic microbiota and its food webs. Patterning in global ozone, global and local climate conditions and annual consumption of light energy in surface-dwelling biota (photosynthesis and other biotic processes) are closely linked together.

Since critical atmospheric thresholds are determined for the worst possible environmental conditions, one can evaluate bromine and oxygen critical thresholds for the springtime polar Arctic environs.

In conventional medicine, symptoms are considered to be a manifestation of the disease. Hence, treatment is given to kill the bacterial life agents causing it or to dampen the symptom of the condition. On the other hand, homeopathy sees the symptoms as a body's attempt to restore the balance, to heal itself. The homeopathic view can be useful for analysis of associated phenomena of the Arctic bromine emissions. Distributed life component is fighting the ozone-oxygen stress till the stabilization of DO in Arctic marine environs. From homeopathy we borrow (1) principle of similarity and (2) principle of minimum dose. In a context of bromine explosion, principle of similarity is likely translated as: trade-off terminal oxidants by means of metabolic emissions. The rate (and the final dosage) of bromine emitted into boundary layer is such that it may able to increase pro (or against) life conditions only slightly. Simultaneously, "training" of the active life component and formation of balanced habitable environments take place.

Because doses of emissions are small, there is no long lasting hostile-life effect or side effect. A remedy starts from the level of the Earth's life (4) downward to the level (1), which is a level of the instance of a life form. Earth's life has an opportunity to exercise its abilities in efficient control of ABL ozone-oxygen; in case of so-called butterfly effect, Earth's life has even a chance to expend its sensitivity.

Changes in environs proceed till the new state of environs is reached, and balance in the habitable environments is set up. Phenomena become non significant or submissive to the phenomena characterizing the environs in their new state. We consider balanced environment as an environment that has maximum production of its Earth's life component for the given external energy, nutrients and other shared resources.

We suggest that the true causes for phenomena have to be determined by the reversing of their appearance order. If so, the significance order of the causes of Arctic bromine explosion is: (1) limited DO in SML (cause which primarily ascends to the paleo marine microbiota), (2) excess of surface ozone in ABL (cause which mostly ascends to the surface-dwelling Earth's type of life).

## § 5.3. CRITICAL THRESHOLDS FOR REACTIVE BROMINE AND OXYGEN SPECIES IN THE EARTH'S SUPERFICIAL MICROBIOTA
**Index: Sensitivity to the reactive oxygen and bromine compounds at the low temperatures**

Life forms at the base levels of the Earth's life compete in efficiency of their metabolic pathways. In Arctic environments, success in this competition is mainly determined by sensitivity to the reactive oxygen (ozone and oxygen) and bromine compounds at the low temperatures.

Co-integration in the Earth's life provide physical basis for a quorum sensing of the Earth as a whole (4). In the pro-life environments spread over the Earth's surface, the evidence of co-integration of the food webs is largely burying down. It is the extreme environments which pay witness to the conjunction of the surface-dwelling microbiota and microbiota in the deep ocean and under earth.

We are specifically interested in the exploration of the extreme environments covering the large surface areas, e. g. the polar Arctic environments. Biotic oxygen production of the Arctic ocean is a seasonal phenomenon. Withstanding productive period, the open ocean-surface serves as a sink for the atmospheric oxygen. At dark and very cold winter, tropospheric ozone remains at high levels due to the stratospheric injections to the ABL. At polar sunrise, the ABL bromine emissions cause to the tropospheric ozone-oxygen conversion and intensification of oxygen fluxes. Even the solar luminosity holds its interannual and annual patterns, the ongoing biotic influence on the atmospheric composition has the potential to initiate substantial climate changes. The ocean life sustaining under the wintertime sea ice/ snow pack is difficult to study directly. Instead we can use the bromine flux marker of the ocean biotic activities. The density and distribution of the bromine flux are an indicative of the interplay between (i) the surface ozone concentrations, sunlit return and (ii) needs of the local photosynthetic marine organisms in dissolved oxygen. Substantial day/ night light variations during the transition season imply that wintertime ozone flux "condenses" near the surface, and the Arctic ABL is buffered by the surface ozone/ dissolved oxygen vapor disequilibrium far from the life-sustaining conditions. Bromine influx is assumed to be an output of snow blowing transport and photochemistry in snow medium, and affected only by the ozone agent and temperature and light conditions of the near-surface atmosphere.

The pivot point in our considerations are local superficial life agents and habitable environments. We would assume that during "transition season" in polar Arctic, local atmosphere-climate system is being restricted to the first 1-2 km of ABL, or even to the micro climate layer in vicinity of halogen-polluted sea ice/ snow pack, we focus on the "material" budget modeling of ozone-oxygen transformation.

Occasion of ozone-oxygen transformation at and near habitable surface is particularly significant if the sufficiently active oxygen-photosynthetic biotic component is absent. For the near to standard conditions, sensitivity to ozone and oxygen in the superficial Earth's type of life is evaluated as 50 ppb of surface ozone and 10-35 % atmospheric oxygen content; sensitivity to the dissolved oxygen is not verified yet.

Though bromine species are minor constituents of the ABL, they efficiently destroy the surface ozone. While alone, halogen and other biotic-originated emissions from the Earth's surface (e.g. sulfur emissions, $SO_2$ photochemistry) couldn't produce an affluence of the atmospheric oxygen column, they can cause abundance of the dissolved oxygen in SML. Indeed background halogen, sulfur and phosphorus ocean emissions into atmosphere is a dominant regulator of global productivity of the superficial oxygen-photosynthetic microbiota and its food webs.

Due to a lack of knowledge, right now we cannot suggest a conceptualization for the atmospheric oxygen metric.

## § 6 TRANSCENDENTAL PHENOMENA
**Index: Sensing of the everyday-world material phenomena and real-life phenomena**

"In natural science the object of investigation is not nature as such, but nature exposed to man's mode of enquiry." (W. Heisenberg)

No matter what we think of the beingness and transcendental life, we can't help but admire wonders of the everyday-world. Admiration of the everyday-world beingness leads to a quest for knowledge and quest for timeless purpose. Transcendental sense of purpose and its serene certainty needs to be borne in transcendental mind that wants to share his quest with others.

In the Earth's type of life, sensible adaptations to the material phenomena are possible due to recognition and memorization processes embedded in the metabolic pathways. Origins of sensible adaptations are followed beyond the physical division of the Earth's system.
At the base levels of the Earth's life hierarchy, sensing is provided by direct molecular signaling; however, molecular signaling often has a limited response. In the large life forms, e.g. human-like life forms, sensing involves many interpretations or responses; thus it is possible to force the effect into many familiar pathways. By means of quorum sensing (4), sensing and real-life perception can be attributed to the Earth's type of life as a whole.

## § 6.1. CONCEPTUALIZATION OF TRANSCENDENTAL PHENOMENA AT THE LEVEL OF THE EARTH'S TYPE LIFE
**Index: Conceptualization of the everyday-world phenomena: Earth's life as a mindset of the active life agents distributed over the Earth's system, Earth's atmosphere is a physical-chemical information environment; role of the atmospheric processing for the functioning of the Earth's life, power law distributions in the distributed networks**

Gödel's Incompleteness Theorem: "Anything you can draw a circle around cannot explain itself without referring to something outside the circle - something you have to assume but cannot prove."

We suggest that superficial life forms are equal observers of the physical reality. We assume that perception processing of the Earth's type of life is what makes all distributed life agents to be equal observers. We suggest a kind of perception, which is placed in the 3+1 Minkowski space, because conformal to our human perception, physical everyday-world is placed in the 3+1 Minkowski physical-chemical space.
Observed physical-chemical reality is a very large Gödel's circle. Unknown dimensions, states and terms of states of universe (and biotic Earth's system) are located outside of it. Associated with sensing of the Earth's life forms, self-reflecting ability of observer is present or seems to be present inside the circle. Supported by material sensing, self-reflective, spiritual life component of observer still refers to the transcendental or a-prior knowledge (Socrates, I. Kant).
Placed in stochastic physical-chemical reality, observer has no means to decide whether factor of influence existed before the influence had been distinguished by self-reflecting perception.
A- prior knowledge is accessible through self-reflecting perception. A-prior knowledge relates to the natural phenomena as to the stochastic phenomena. The outcomes of stochastic phenomena may and may not result in a specific pattern of outcomes. The enigma is that some natural phenomena often result in patterns of outcomes that do not appear to be the outcomes of random chance events. In historical practice, modified forms of the diverse natural forces are recorded and conceptualized by preceptors of a human life form; upon fortunate conditions, some preceptors become facilitators of the material development of a human civilized life form.

Of the essence, observed natural phenomena are a function of a number of actual trials of the events. Diverse driving forces govern the interminable development of the Earth's system and development of atmosphere, but none of these forces confines to one particular conceptualization. Any particular conceptualization refers to the regular patterns in the past-time outcomes, but patterns do not rule over random chance events. Conceptualization of the everyday-world phenomena is improbable without conceptualization of the real-life world. Success of conceptualization of the real-life world depends on our understanding and interpretation of the origin(s) and driving forces of universe.

In order to identify a framework for atmospheric modeling, we suggest a first exegetical conceptualization of the Earth's atmosphere-climate system and its control factors:

(1) The real-life/ everyday-world Earth's atmosphere-climate system is a stochastic system.

Despite all, a majority of computational models of the global troposphere is deterministic; models imply the fixed numerical schemes of the reversible evolution equations, that would supply the unique and stable working solution (e.g. sets of solutions for DaisyWorld) for the outcomes of the several specific patterns. It is evident that the evolution equations exist only in the anthropocentric rational perception. Moreover, deterministic atmospheric models always have a number of "screwed" parameters and are not actually designed to deal with forward-in-time local climate changes or the global climate transitions. Being subjects to the logical laws, we always try to introduce our reasons and concerns to justify new conceptions. Even though, there is no way to prove that human interpretations are based the matter connections only, or to prove that development of the biotic Earth has its source in a dispirited matter.

(2) The assumption about physical forcing - confinement of the control factors to the 3+1 Minkowski space is not quite obvious. Although the contemporary implementations of these control factors <which are limited to the "conventional math"> provide the reasonable picture of some observed phenomena, the general consistency of the real-life out- of- time "*statistical*" driving forces has never been questioned. For example, in view of the continual appearance of new connections ( and, consequently, new parameterizations) during the development of the biotic Earth's system, Earth's life forms must extend their *statistical sustainability*. There is no way to prove that *statistical sustainability* is based temporal everyday-world connections only.

(3) Through self-reflecting spiritual perception, something outside of a circle of the physical-chemical medium persistently interacts with unity of the Earth life forms to promote purpose of existence and development of the Earth's life.

(4) The everyday-world atmospheric systems are systems undergoing the irreversible and reversible state transitions. Atmospheric systems similar to the Earth's troposphere, the Earth's stratosphere, the planetary atmospheres, are open for matter and energy exchange. Geological records point out on the variations in the biogeochemical cycling. Present atmospheric parameters also signal on the variations (carbon and hydrocarbon chemical substances, etc) in the biogeochemical cycling upon the interminable partaking of the biotic Earth. Complexity of biogeochemical cycling in everyday-world biotic Earth is intrinsically linked to the multiplicity of time scales and fractional topology. Collection of the diverse life forms who communicate with each other only through the physical-chemical medium, can't produce the collective, hierarchical judgments < true vector of development>. It is an original installation of the transcendental life component perceived by all, which is necessary for communication beyond 3+1 space.

(5) From the spiritual perception of imaginative human individuals, we learn that reliability of transcendental life not known in any justification other than its own self-reflection. However, once perceived, a-prior knowledge is able to be absorbed by multiple individuals indifferent to their age and historical age, sex and professional everyday skills. Emerging knowledge derived from experience and believes is organized by a-prior established knowledge.

During the process of abstraction and modeling, Earth's subsystems are usually reduced to the simpler form of event- driven systems working as Finite State Machines (FSMs). FSMs are machines that tend to return relatively fast to the previous state or transfer to the new state upon the regular disturbances, spatial and temporal. In case of the atmosphere-climate system, the stable states of atmosphere are seasons of the cosmic cycle. For instance, balanced by the diversity of natural factors ([3], [6]) that provide four tropospheric near-equilibriums, the Earth's troposphere has four distinct seasons in the middle latitudes. For the high latitudes, there is common to speak about daylight and nighttime seasons and transition periods between them.

It is important to mention that mechanistic approach to the natural phenomena and regulation of Earth's system, as directing by dispirited matter, is relatively new. It never was practicing by the originators of the modern science (T. Kuhn, 1962). The truth is that in the real-life world, everything is connected with everything, and everything has its structured, hierarchal meaning. We cannot approach the understanding of the natural phenomena by excluding the observer place of hierarchy, whether a human observer or the Earth's life place of hierarchy. Because the Earth's life is presented by the mindset of active life agents distributed over the Earth's system, therefore, it is appropriate to treat < to conceptualize> natural phenomena and natural systems as processing over the complex networks of the active agents. Through their association with life forms, life agents also communicate by means of the physical-chemical medium. At the seasonal physical-chemical-dynamical conditions, such complex network has a stable rank linkage expressed in physical-chemical terms. The corresponding rank linkage is only one of many abstract expressions. Apparently, such rank linkage is defined by the energy-matter transformations available for the life control. The pattern rules of linkage are such that they guarantee the system's seasonal equilibrium in the interest of the Earth's life and its development.

## § 6.2. LIFE AGENTS IN THE EARTH'S MATERIAL DOMAIN
**Index: Transcendental life and development of the biotic Earth, mental (mind) activities of a human observer**

"To put it boldly, science is the attempt at the posterior reconstruction of existence by the process of conceptualization", A. Einstein

Abstraction, or conceptualization, is associated with a human rational and irrational perception, which involves phenomena of the transcendental self-reflection. Through all its increasing presence on the planet, life regulates the development of the Earth's planetary system. Even though it seems plausible at the moment different from the origin, life and organic Earth's type of life have a strong connection with each other.

Subsidiary theme of this research is transcendental influence on the development of the physical division and on the Earth's life perception toward enforced development and the purpose of existence.

Transcendental life in a human life form takes presence in what we call sometimes mental activities and attitudes. Although it was always admitted, by a part of human population, that self-reflecting perception in superficial large life forms cannot be deduced to the material phenomena, the mainstream scientific research of the last decades completely ignores impact of the transcendental life on the biogeochemical processing and, consequently, on the development of the Earth's system.

Arisen from the extermination of the agnostic views, contemporary bulk science considers Earth's type of life as assemblage of chemical species in physical-chemical domain. Indeed, due to the terms of sensing in the Earth's life forms, observations and self-observations are commonly restricted to the everyday-world phenomena. Yet, while at the atomic and molecular scales, material phenomena can be described by chemical science, at particle's scales, chemical abstraction of phenomena is stripped away. Energy and matter are equivalent: matter can be turned into energy and energy into matter. Being more or less agnostic in relation to the physical-chemical phenomena, facilitators of the modern science extensively related their achievements in the construction of conceptions to the spiritual moments of enlightenment, dreaming and associative sensing. Though their personal faiths differ one from another, founders of the modern science were very creative people and not only in doing science. Being less creative <spiritual>, at least in terms of the systematic nature exploration, conventional bulk-scientists discredit transcendental beliefs on the constant basis.

## § 6.3. HUMAN SENSING, REASONING AND PARTAKING
**Index: Probabilistic character of natural (earthy) phenomena, interminability of the Earth's development, temporal limits on the organic Earth's life forms**

"The human being himself," W. Goethe insisted, "to the extent that he makes sound use of his senses, is the most exact physical apparatus that can exist."

If the concept of infinite dichotomy is to be defined by means of verification, the question arises how the means of verification to be determined: are they a matter of conventional choice or originally required by life and have been proceeded from the early stage of the biotic Earth's development. In fact, the only means of processing that is consensual by the all human observers is a time property of the physical world. While modern science uses light propagation for the arbitrary time measurements, in a past, time property was measured by the sunrises and sunsets and other periodical phenomena. Earthy phenomena are the probabilistic phenomena. We stipulate that only the Earth's life and the Earth's system are bounded to the property of time, and life is not bounded. If this is a case, the probabilistic decision about whether outcome of the chain of events will be pro-life – pro-Earth's life outcome is on the responsibility of the transcendental life. According to the human reasoning and human-like sensing, kind of processes that one chooses is a kind of processes concerning which one knows something certain. On the inductive grounds we conclude that transcendental life should play such a central and synchronization role in the development of the Earth's life and the Earth's system, that in order to succeed in development into the highest form of themselves, the Earth's life forms should follow (and include in their metabolism) the empirical processes of its predecessors and the sensing knowledge that transcendental life passed to them.

How the transcendental life passes the knowledge to the Earth's life agents? A-prior knowledge is acquired at different levels of sensing, from the level of quorum sensing (4) to the level (1). It is merely interesting how the knowledge is passed to the individual life agent. In spite of the huge difference in the life expectancy, life style and experience, the metaphysical conception of transcendental life was familiar to human-like life forms before the advent of epoch of human civilization. Discovery of transcendental life takes place through its actual communication with Earth's life forms at the all levels of sensing, and especially, at the level of a quorum sensing.

One of the proliferated features of the Earth's large life forms is ability and necessity to sleep. Sleeping phase is confirmed in many invertebrates and in all vertebrates, from insects to fish and mammals. Sleeping is very different from the conservation phase. Sleeping is very important, for humans it takes more than half of a lifetime. As observed, complexity of behavior of healthy individual of a human life form depends on the quality of sleeping phase.
For vertebrates, complexity of behavior doesn't seem to depend on the brain's size or its fine segmentation, but depends on sleep. Young mammals sleep more than adult mammals of the same kind. Prenatal womb sleep is essential condition for the development of the healthy organism.
Sleep has vital functions both for life and the Earth's life. The genetic machinery of ontogenetic growth doesn't cancel the most probable reason for prenatal fetal sleep-and-dream - life control over the mental sensing and the physical processes of differentiation, growth and healing. Deprivation of sleep always ends in death, which is preceded by mental and psychological destructions. Human ancestors believed that during the sleep soul got less connected to a body. The ancient Jews believed that the proper, natural way to die is during the sleep, when the soul-life voluntary abandons the body-life form, goes to a source and joins other souls. During the dream-sleep cycle, mammals see dreams. Many types of dreaming are known. Long ago, special dreaming experience was considered as a guidance experience similar to meditation experience. People often try to understand meaning of their dreams and pursue a guidance. Particular imaginative and creative individuals - composers, poets, mathematicians and others can experience dreaming during their regular thinking activities.
Even we know that sleep is absolutely necessary and a concept of soul is ordinary in all human cultures, we don't know what a soul is and what happens to a soul when body is sleeping.

Interpretation of thinking and dreaming experience is very individual and heavily depends on the everyday-world surrounding and circumstances of the individual. Any interpretation is relative; it would suffer distortion and loss of original meaning being applied to the finite temporal phenomena.

However, it is important to notice that some dreaming interpretations and general concepts produced by preceptors-visionaries become accepted by significant part of human population and utilized in the development of a human civilized life form. We know very little how and why the choice of general concepts and beliefs that become a part of culture, science and technology is made, because it seems to be a natural choice. It is like the creative spark of transcendental consciousness within people linking everyone to the creative mind of a universal life thus making us co-creators of the physical everyday-world and de-restrict our "exakte sinnliche Phantasia of Urphanomën" (Goethe)    .

# § 7 APPENDIX

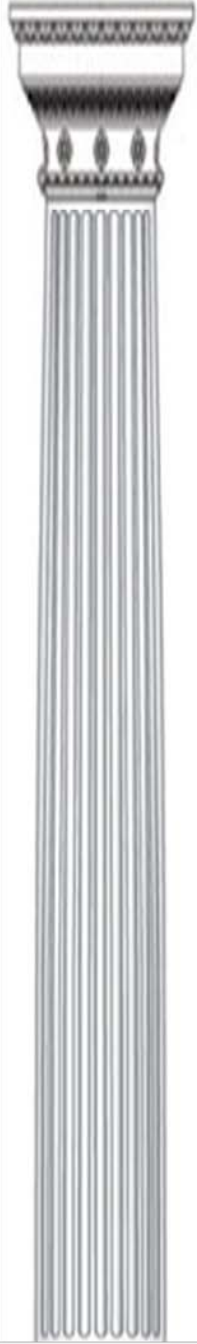

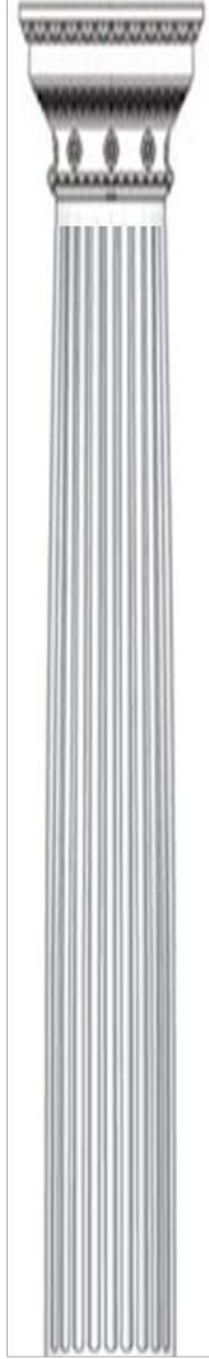

### PLATO IN *TIMAEUS*

Plato reveals a lot of transcendental mind in his writings. In the *Timaeus*, Plato writes about soul that has a view (of the heavens) in a prior state of being. He also shares his perception of the physical world - "in the same way space or matter is neither earth nor fire nor air nor water, but an invisible and formless being which receives all things, and in an incomprehensible manner partakes of the intelligible". Plato describes nourishment to the animal body and destruction of organic body as in motion (development). Plato explains being as unity, the most general symbol of relation. In Plato's *Timaeus*, the world of Gaea is treated as a living Gaea, a unity of living things.

For the many years Plato has been interpreted through the Greek philosophy and polytheistic religion (gnostics, modern science). He also has been interpreted through the Jewish faith and philosophy (Phylo, cabbalists, masons). The repetitive discovery of a new (Leibnitz, Goethe, Lovelock & Margulis) meaning of Plato's writings signs beginning of reformation of the rational perception toward the quorum sensing of the Earth's life as a whole

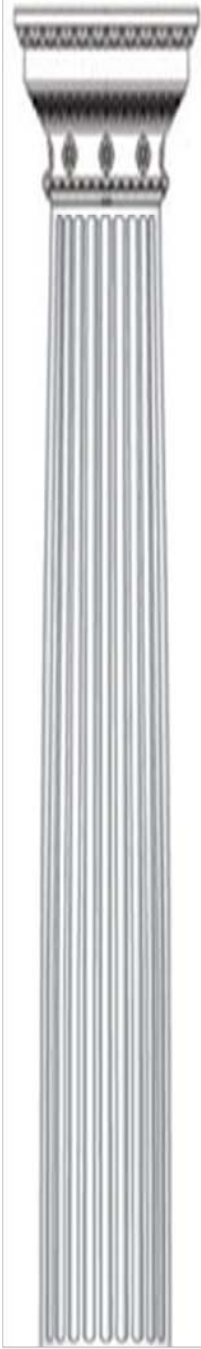

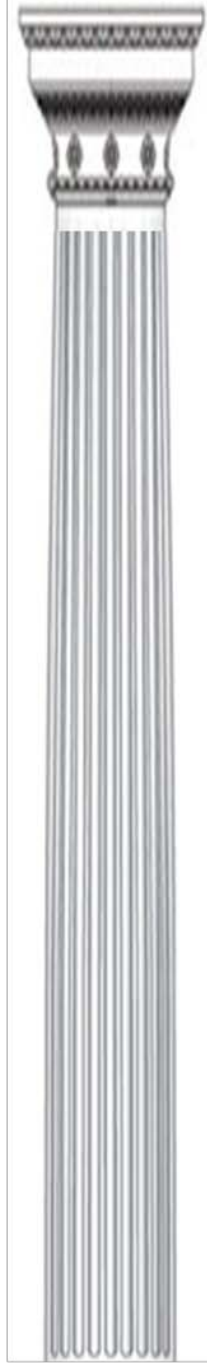

## GOTTFRIED VON LEIBNITZ IN *PROTOGAEA*

**§2. The globe of the earth was at first a regular form, and has hardened from liquid, light or fire being the motive cause.**

**§2.Pr p14 Consequently, the hard coating that we now see was made later; at one time, however, it must have been liquid. Now fluidity itself arises from internal motion, and as if from a degree of heat, which experiments prove. For even water turns to ice because of reduced heat; while on the contrary corroding liquids, being strong because of a hidden motion, are hardened with difficulty. Moreover, heat or internal motion comes from fire or from light, that is, from a very rarefied, penetrating spirit. And thus we have arrived at the motive cause which Sacred History also takes as the beginning of Cosmogony.**

**§6 From where did the water come that covers the Earth and what happened to it? And on the various causes of the flood. Just as fire in the beginning seized everything, before the light was separated from the darkness, so it is supposed that after the extinction of this fire everything was submerged by the waters. ...nothing appears more plausible than that we might believe that the vault of the Earth ruptured where it was held up by weaker supports, and a huge mass collapsed into the formerly enclosed sea below, leaving peaks exposed. So the waters, forced out of caves, overflowed above the mountains, until, having found a new way into the underworld when the floodgates of the inner Earth were broken open, they left anew whatever now appears as dry land.**

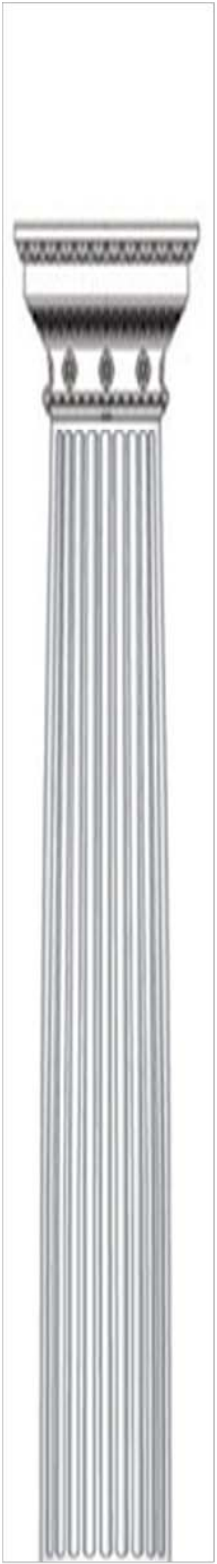

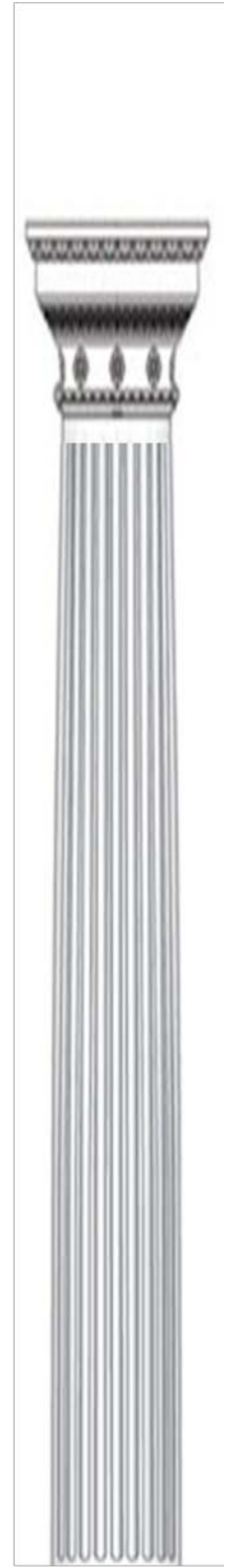

### GOETHE IN *PARABASE*

Freudig war vor vielen Jahren,
Eifrig so der Geist bestrebt,
Zu erforschen, zu erfahren,
Wie Natur im Schaffen lebt.
Und es ist das ewig Eine,
Das sich vielfach offenbart:
Klein das Große, groß das Kleine,
Alles nach der eignen Art;
Immer wechselnd, fest sich haltend,
Nah und fern und fern und nah,
So gestaltend, umgestaltend —
Zum Erstaunen bin ich da.

Joyously, so long ago,
My eager mind did strive
To study and discover
Nature in her works alive:
She, the everlasting Oneness
In the manyness divined.
Big minuteness, tiny bigness,
All according to its kind,
Ever changing, ever constant
Near and far, far and near,
Shaping and reshaping . . .
Why but to wonder am I here!

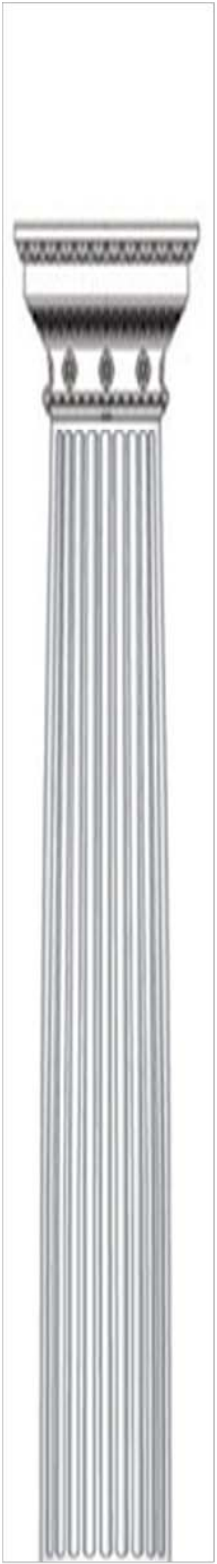
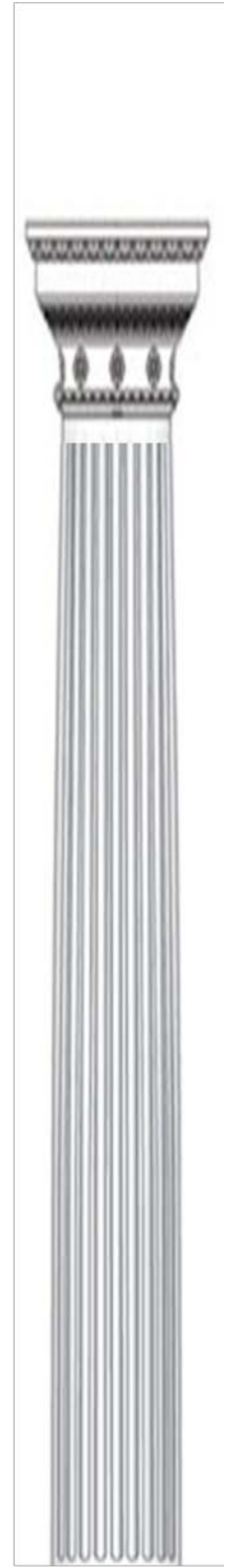

**FRANZ MARC IN *CORRESPONDENCE***

**German artist- expressionist Franz Marc  wrote in his letters ( 1910-1914) about  personal sensing experience of a wonder and beauty of the Earth:**

**I am trying to increase sensitivity to the organic rhythm of all things, trying to achieve a pantheistic empathy with throbbing and racing of the blood in nature, in the trees, in the animals, in the air — trying to make this into a picture with new movements...**

**We no longer cling to the image of nature, but destroy it, in order to expose the powerful laws that hold sway behind the beauty of appearances...**

**I cannot detach myself from my thoughts and dreams, nor do I even want to .. I am beginning to look more and more behind or, to put it better, through thing, a behind that things tend to conceal with their appearance**

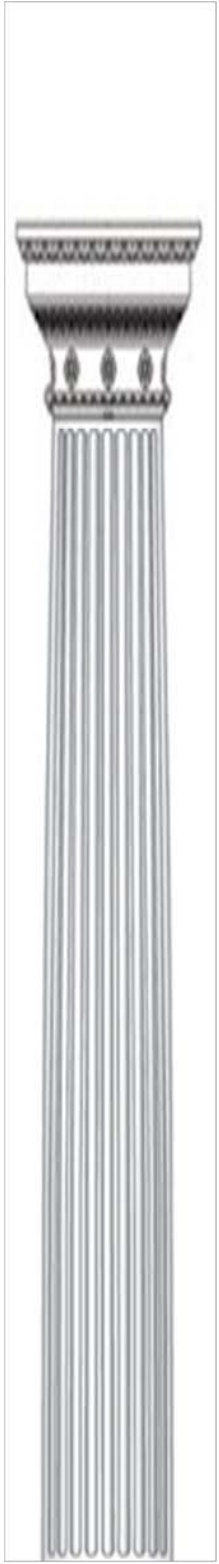
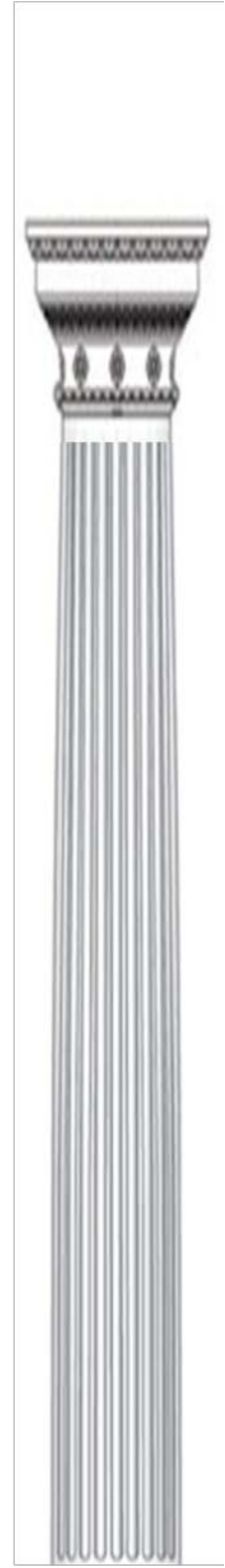

## LYNN MARGULIS ABOUT
### *GAIA AND SYMBIOGENESIS*

**Symbiosis is a physical association between organisms, the living together of organisms of different species in the same place at the same time... I contend that symbiogenesis is the result of long-term living together — staying together, especially involving microbes- -and that it's the major <evolutionary> innovator in all lineages of larger nonbacterial organisms...**

**James Lovelock believed that the gases in the atmosphere were biological. These gases were far too abundant in the atmosphere to be formed by chemical and physical processes alone. He argued that the atmosphere was a physiological and not just a chemical system...**

**The Gaia hypothesis states that the temperature of the planet, the oxidation state and other chemistry of all of the gases of the lower atmosphere (except helium, argon, and other nonreactive ones) are produced and maintained by the sum of life**

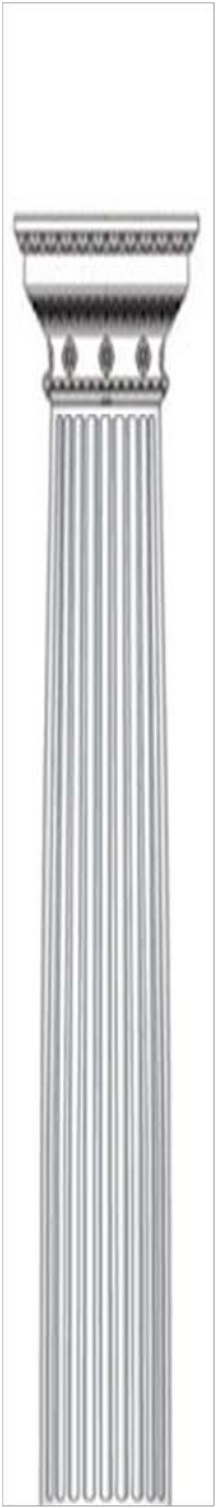
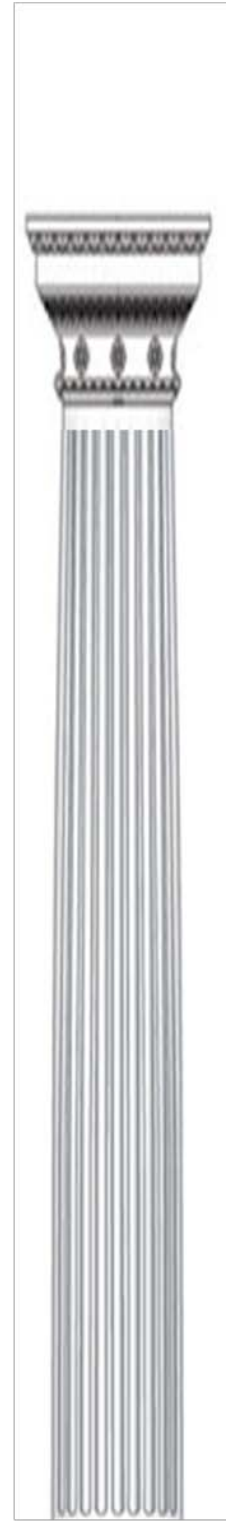

### Plato

**Thinking is speaking to the soul**

### Pierre Teilhard de Chardin

**We are not human beings having a
spiritual experience.
We are spiritual beings having a human
experience.**

### Henry Austin Dobson, In After Days

In after days when grasses high
O'ertop the stone where I shall lie,
Though ill or well the world adjust
My slender claim to honored dust,
I shall not question nor reply.
I shall not see the morning sky;
I shall not hear the night-wind sigh;
I shall be mute, as all men must
In after days!
But yet, now living, fain would I
That some one then should testify,
Saying-"He held his pen in trust
To Art, not serving shame or lust."
Will none?-Then let my memory die
In after days!

# REFERENCES

1. Bashkin, V., (2003), *Modern Biogeochemistry*, Kluwer academic publishing
2. Fabrikant A.,   et  d., (2003), On a network creation game. In Proceedings of ACM Symposium on Principles of Distributed Systems, pp. 247–351
3. Galaktionov, V. et al., (1997), *Aircraft observations of ozone in the Arctic troposphere in April 1994,* Atmospheric Research Volume 44, Issues 1-2, pp. 191-198
4. von Glasow, R. et al., (2009), The effects of volcanic eruptions on atmospheric chemistry, in Halogens in Volcanic Systems and Their Environmental Impacts, Chemical Geology, 263 (1-4), pp. 131-142
5. Gabriel, A. et al., (1999), *The Tropopause in a 2D Circulation Model*, Journal of the Atmospheric Sciences; 56: 4059-4068
6. Grimes, H.D. et al., (1983), *Ozone degrades into hydroxyl radical under physiological conditions: a spin trapping study*, Plant Physiol 72: 1016–1020
7. Hazen ,R. M. et al., (2002), *High pressure and the origin of life,* J. Phys.: Condens. Matter 14: 11489–11494
8. Herbert, D. et al. ( 1999), Effects of plant growth characteristics on biogeochemistry and community composition in a changing climate,   Ecosystems 2:367-382
9. Hines, M. et al., (1998), Rapid Consumption of Low Concentrations of Methyl Bromide by Soil Bacteria, Appl. Environ. Microbiol., 64(5): 1864–1870
10. Iudin, M., (2007), Study of fundamental physical principles in atmospheric modeling based on identification of atmosphere -climate control factors, part 1: *Bromine Explosion At The Polar Arctic Sunrise*, arXiv:0712.2723v3
11. Iudin, M., (2008), Study of fundamental physical principles in atmospheric modeling based on identification of atmosphere - climate control factors, part 2: *Gaia Paradigm: A Biotic Origin Of The Polar Sunrise Arctic Bromine Explosion*, arXiv:0812.4797v3
12. Jordan,   A.,  (2003), *Volcanic formation of halogenated organic compounds in   The Handbook of Environmental Chemistry*, vol. 3. Springer-Verlag, Berlin,Heidelberg, pp.121–139
13. Jørgensen, B. et al., (1986), Transition from anoxygenic to oxygenic photosynthesis in a Microcoleus chthonoplastes cyanobacterial, Appl. Environ. Microbiol. mat. 51:408–417
14. Levine, J. & Augustsson, T., (1985), *The photochemistry of biogenic gases in the early and present atmosphere,* Origins of Life and Evolution of Biospheres, Springer
15. Lovelock, J. E ., (1972), *Gaia as seen through the atmosphere*, Atmospheric Environment 6 (8): 579–580
16. Margulis, L., (1999), *Symbiotic Planet: A New Look At Evolution*, Houston: Basic Book
17. Myerson, R. B., (1977), *Graphs and cooperative game,* Mathematics of Operations Research, 2(3)
18. Myerson, R. B., (1991), *Game Theory: Analysis of Conflict*, Harvard University Press
19. Newman M., (2003), *The structure and function of complex networks*, SIAM Review, 45(2): 167—256
20. Neilson, A. H., (2003*), Biological Effects and Biosynthesis of Brominated Metabolites*, The Handbook of Environmental Chemistry Vol. 3, Part R, pp. 75–204

21. Parker E., (1979), Cosmical Magnetic Fields: Their Origin and their Activity, Oxford University Press
22. Perelman, A. I. , (1989)   *Geokhimiia* (*geochemistry*) *epigeneticheskikh protsessov*, Moscow, Visshaya Shkola, English edition New York : Plenum Press
23. Pikovsky, A. et al., (2001), *Synchronization—A Unified Approach to Nonlinear Science* Cambridge University Press, Cambridge
24. Pikovsky, A., (1984), in *Nonlinear and Turbulent Processes in Physics*, Harwood Academic, Singapore, edited by R. Z. Sagdeev,   pp. 1601–1604.
25. Pomeroy, L.R, (1997), Primary production in the Arctic Ocean estimated from dissolved oxygen, Journal of Marine Systems 10, pp. 1–8
26. Pollack, H. N. et al., (1993), *Heat Flow from the Earth's Interior: Analysis of the Global Data Set*, Rev. Geophys., 31(3), pp. 267–280
27. Rastetter, E. B. et al., (2005), Carbon Sequestration in Terrestrial Ecosystems Under Elevated $CO_2$ and Temperature:   Role of Dissolved Organic Nitrogen Loss, Ecological Applications 15:71-86
28. Rastetter, E. et al., (2001), Resource Optimization and Symbiotic Nitrogen Fixation, Ecosystems 4:369-388
29. Rastetter, E.et al, (1997), Responses of N-limited ecosystems to increased $CO_2$: A balanced-nutrition, coupled-element-cycles model,   Ecological Applications 7:444-460
30. Robock, A., ( 2000), *Volcanic Eruptions and Climate,* Rev. of Geophysics, v. 38(2): 191-219, DOI:10.1029/1998RG000054
31. Shi-Ying Lin et al., (2001), *Hydrogen Production from Hydrocarbon by Integration of Water−Carbon Reaction and Carbon Dioxide Removal* (HyPr−RING Method), Thermal Energy and Combustion Engineering Department, National Institute for Resources and Environment, Japan   Energy Fuels*,* 15 (2), pp. 339–343 DOI: 10.1021/ef000089u
32. Tardos, E. & Wexler, T., (2007), *Network Formation Games* in Algorithmic Game Theory, Cambridge University
33. *The Limits of Organic Life in Planetary Systems,* SSB ( Space Studies Board) and BLS (Board on Life Sciences), 2007, http://www.nap.edu/catalog/11919.html
34. Usoskin, I. G. et al., (2007), *Grand minima and maxima of solar activity: new observational constraints***,** A&A 471(1), pp. 301-309, DOI: 10.1051/0004-6361:20077704
35. Vingarzan R., (2004), *A review of surface ozone background levels and trends,* Atmospheric Environment 38: 3431–3442